\documentclass{article}
\usepackage{graphicx} 
\usepackage[a4paper, total={6.1in, 10in}]{geometry}
\usepackage[a4paper]{geometry}

\usepackage{geometry}

\usepackage{subfiles}

\usepackage{natbib}
\setcitestyle{author,open={(},close={)}} 

\usepackage{array}
\usepackage{booktabs}
\usepackage{amsmath}
\usepackage{amssymb}
\usepackage{float}
\usepackage{hyperref} 
\usepackage{gensymb}
\usepackage{mathtools}
\usepackage{wasysym}
\usepackage{mhchem}

\usepackage{authblk}

\usepackage{commath}

\usepackage{xcolor}

\makeatletter
\newcommand*{\centerfloat}{%
  \parindent \z@
  \leftskip \z@ \@plus 1fil \@minus \textwidth
  \rightskip\leftskip
  \parfillskip \z@skip}
\makeatother

\title{Hybrid machine learning data assimilation for marine biogeochemistry}

\author[1,2]{Ieuan Higgs}
\author[1,2]{Ross Bannister}
\author[2,3]{Jozef Sk\'akala}
\author[1,4]{Alberto Carrassi}
\author[5]{Stefano Ciavatta}
\affil[1]{Department of Meteorology, University of Reading, UK}
\affil[2]{National Centre for Earth Observation, UK}
\affil[3]{Plymouth Marine Laboratory, UK}
\affil[4]{Department of Physics and Astronomy ``Augusto Righi'', University of Bologna, IT}
\affil[5]{Mercator Ocean International, FR}

\begin{document}

\maketitle
\begin{abstract}
Marine biogeochemistry models are critical for forecasting, as well as estimating ecosystem responses to climate change and human activities. Data assimilation (DA) improves these models by aligning them with real-world observations, but marine biogeochemistry DA faces challenges due to model complexity, strong nonlinearity, and sparse, uncertain observations. Existing DA methods applied to marine biogeochemistry struggle to update unobserved variables effectively, while ensemble-based methods are computationally too expensive for high-complexity marine biogeochemistry models. This study demonstrates how machine learning (ML) can improve marine biogeochemistry DA by learning statistical relationships between observed and unobserved variables. We integrate ML-driven balancing schemes into a 1D prototype of a system used to forecast marine biogeochemistry in the North-West European Shelf seas. ML is applied to predict (i) state-dependent correlations from free-run ensembles and (ii), in an ``end-to-end'' fashion, analysis increments from an Ensemble Kalman Filter. Our results show that ML significantly enhances updates for previously not-updated variables when compared to univariate schemes akin to those used operationally. Furthermore, ML models exhibit moderate transferability to new locations, a crucial step toward scaling these methods to 3D operational systems. We conclude that ML offers a clear pathway to overcome current computational bottlenecks in marine biogeochemistry DA and that refining transferability, optimizing training data sampling, and evaluating scalability for large-scale marine forecasting, should be future research priorities.
\end{abstract}

\section{Introduction}
\label{sec:background}
Marine biogeochemistry (BGC) modelling is an essential tool for understanding global marine elemental cycles (e.g., for carbon and nitrogen), as well as for understanding the response of marine ecosystems to a range of human and climate pressures \citep{heinze2013modeling, ford2018marine, fennel2022ocean}. These pressures include ocean acidification, marine heat waves, and nutrient pollution, and lead to a range of consequences, such as deoxygenation, toxic algal blooms and biodiversity loss \citep{doney2009ocean, smith2009eutrophication,  schmidtko2017decline, frolicher2018emerging, fennel2019biogeochemical, gobler2020climate}. Marine BGC modelling could then support management, policy and planning across a wide range of temporal scales. Marine BGC models are often constrained by the available observations through data assimilation (DA) \citep{ford2018marine, fennel2019advancing}, providing both multi-decadal reanalyses of past ecosystem trends and variability, as well as short-term operational forecasts (on the scale up to 5-10 days). Such operational forecasts are run by marine forecasting centres in many countries, e.g., by the Copernicus Marine Service in Europe covering the global ocean and all the major European seas \citep{le2019observation}. 

However, marine BGC DA faces multiple specific challenges \citep{dowd2014statistical, ford2018marine, fennel2019advancing}, compared to assimilation of ocean physics observations in marine models. Marine BGC models are typically more complex than physical models (a pelagic model can have tens of variables and hundreds of parameters), they are highly non-linear and relatively poorly constrained (e.g., having highly uncertain parameters) when compared to ocean physics models. 
Furthermore, marine BGC observations are even fewer, sparser, and more uncertain than physics observations. This brings several specific challenges for marine BGC DA, one of those being the need for multivariate DA, where a large portion of the marine BGC model state variables is updated by observations of only a small fraction of the model variables. In the context of operational marine BGC forecasting, these observations are typically satellite ocean colour-derived chlorophyll \citep{fennel2019advancing, groom2019satellite}, with assimilation of BGC-Argo observations (including chlorophyll, nitrate and oxygen) in open ocean waters recently implemented in state-of-the-art operational systems \citep{cossarini2019towards, teruzzi2021deep}. 
Other products, such as optical variables \citep{shulman2013impact, ciavatta2014assimilation, jones2016use, gregg2017simulating, skakala2020improved}, 
size-class chlorophyll \citep{ciavatta2018assimilation, ciavatta2019ecoregions, skakala2018assimilation, pradhan2020global} 
and other types of in situ data, such as chlorophyll, oxygen and nutrients from gliders \citep{skakala2021towards}, are assimilated in reanalyses, or research and development (R\&D) versions of the operational systems. For a broader range of marine BGC DA work beyond operational applications, see many other references, e.g., \cite{simon2012gaussian, shulman2013impact, gehlen2015building, simon2015experiences}.

Different DA systems are used across marine BGC forecasting centres, including variational \citep{ford2012assimilating, song2016data, skakala2018assimilation, coppini2021copernicus}, Singular Evolutive Extended Kalman filter (SEEK) \citep{gutknecht2019modelling, ciliberti2021monitoring} and Ensemble Kalman Filter (EnKF) \citep{bertino2021arctic} -based methods. Although ensemble methods (e.g., EnKF) are appealing for their capability to provide uncertainty quantification and cross-covariances, the more complex marine BGC models such as the European Regional Seas Ecosystem Model (ERSEM) \citep{butenschon2016ersem} or the Biogeochemical Flux Model (BFM) \citep{cossarini2017development}, currently rely on variational methods, as running a sufficiently large ensemble in the day-to-day operational forecasting context can be computationally prohibitive. Moreover, for such complex models, variational methods update only a very limited number of unobserved variables, typically using very simple balancing principles based on the simulated structure and stoichiometry of the phytoplankton community \citep{teruzzi2014a3dvarassim, skakala2018assimilation}. We will call such systems with certain approximations ``univariate’’, and systems that update (nearly) all model state variables as a direct result of DA ``multivariate’’. The multivariate updates can happen in the DA step, through ensemble-informed background covariances (as in the EnKF), or, 
through balancing schemes, such as the scheme of \cite{hemmings2008ocean} based on nitrogen mass conservation applied to Nutrient-Phytoplankton-Zooplankton-Detritus models \citep{hemmings2008ocean, ford2012assimilating}).
However, whenever such multivariate schemes were applied to highly complex marine BGC models (in reanalyses, or R\&D), the improvement on non-observed variables was typically marginal, with several variables often systematically degraded by DA (e.g., \cite{ciavatta2016decadal, ciavatta2018assimilation}). This provides a warning on the use of incorrect assumptions in multivariate balancing schemes or in the EnKFs, and the need for better DA and/or ensemble design.

The field of machine learning (ML) has developed rapidly during the past few decades, and has seemingly found function across every level of science and culture, due to the increasing size and availability of datasets and computational power, together with the continued development of algorithms and theory \citep{jordan2015machine, sonnewald2021bridging}. Within Earth sciences, the flexibility of ML paradigms has allowed its use in a huge variety of applications \citep{reichstein2019deep}, including extensive use in physical ocean modelling \citep{van2007fast, nowack2018using, kochkov2021machine}. However, using ML for marine BGC models is comparatively infrequent, with the most common examples found in parameter estimation \citep{mattern2012estimating, leeds2013modeling, mattern2014periodic, schartau2017reviews}.  There are only relatively few applications outside of this domain such as using a statistical emulator to quantify uncertainty \citep{mattern2013sensitivity} and the prediction of hypoxia in shelf sea environments \citep{skakala2023future}.

In this work, we investigate the capability of ML to learn the hidden, non-linear, and complex relations between BGC variables and to use the learned functions within a DA scheme or to fully substitute it. Thus, we are not attempting to emulate or improve BGC models via ML, but instead use ML to improve DA, and specifically to cope with the challenging problem of propagating information from observed to unobserved variables. The main goal of this approach is to introduce multivariate DA into the system, whilst benefiting from the relatively low computational cost of ML.
This study falls within a stream of research aimed at building suitable hybrid ML-DA schemes \citep[see][and references therein]{buizza2022data, cheng2023review}, and, to our knowledge, it is the first such attempt in the context of marine BGC.

We first use ML to learn flow-dependent correlations that are needed within a DA update step. This amounts to a merge of DA and ML, whereby the latter is used to accomplish a task within the DA process.
We demonstrate that such an ML-based multivariate DA is efficient and accurate. As long as enough suitable data are available for training, ML is able to learn and map complex non-linear functions for propagating the information from observed to unobserved portions of the system's state. 

In a second configuration, instead of merging DA and ML, the former is used to produce a training dataset from which ML learns the full DA step, in an ``end-to-end'' fashion \citep{barth2020dincae, fablet2021learning}. Here, the ML task is that of DA as a whole, i.e., given the background state and observations, return the analysis increments to the background for unobserved variables. As mentioned above, we do not intend substituting/improving the BGM model, and our end-to-end learning focuses only on learning the instantaneous DA updates while using the given BGC model to issue the forecasts. Efficient end-to-end learning of the EnKF analysis in chaotic systems has been recently proven by \cite{bocquet2024accurate}.

Specifically, we intend to answer the following questions: (a) Can we make improvements to the existing univariate scheme by updating a limited set of additional variables with an ML model to predict correlations or analysis increments? (b) Can these ML models be extended to effectively update all unobserved pelagic variables? (c) Is the ML model transferable to a new location after being trained on some other location? 

Our work has a potentially important application within the North-West European Shelf (NWES) operational DA system to which it is tailored. Yet we will discuss its generalization to other comparable systems, applied to spatial domains with similar type of marine BGC dynamics. Based on the transferability of the ML model, we speculate whether it is feasible to use the ML model trained in 1D on a 3D domain and propose a methodology for doing so.

The paper is structured as follows. We first give, in Sect.~\ref{sec:model_and_da_setups}, details on the 1D physical model, the BGC model and the configuration used. Also, we establish the setups for the DA workflow, describing the reference univariate scheme (RUS), and the use of the EnKF. 
Then, in Sect.~\ref{sec:new_schemes}, we outline the two ML approaches explored in this work.
We also give detail on the ML architecture and climatological statistics.
Next, in Sect.~\ref{sec:discussion}, we present and discuss our results for: updating nitrate only; updating the entire set of pelagic BGC surface variables; and testing the transferability of the ML model to a new location with different BGC behaviour.
In Sect. 5, we draw concluding remarks, summarise the key findings and discuss future work.

\section{Model and data assimilation setups for biogeochemistry}
\label{sec:model_and_da_setups}
\subsection{Physical model: GOTM}
The Generalised Ocean Turbulence Model (GOTM) \citep{bolding1999gotm} is a 1D water column model for studying hydrodynamic and biogeochemical processes when coupled to a biogeochemical model, in marine and limnic waters. It provides a sufficient balance between realism and computational cost, posing as an ideal candidate for studying new DA schemes in realistic scenarios.
GOTM can be used as a stand-alone model for studying dynamics of boundary layers in natural waters, having hydrodynamic applications in investigations of air-sea fluxes \citep{vagle2010upper}, surface mixed-layer dynamics \citep{sonntag2011phytoplankton}, dynamics of bottom boundary layers with or without sediment transport \citep{umlauf2011diapycnal, falchetti2010nearshore}, and estuarine and coastal dynamics \citep{burchard2009combined}. 

\subsection{Biogeochemical model: ERSEM}
ERSEM \citep{baretta1995european, butenschon2016ersem} is a marine biogeochemistry model that simulates lower trophic levels of the ocean ecosystem, including plankton and benthic fauna \citep{blackford1997analysis}, see Table~\ref{tab:ref_table}. 
The model divides phytoplankton into four functional types based on size: picophytoplankton, nanophytoplankton, microphytoplankton and diatoms \citep{baretta1995european}. ERSEM uses variable stoichiometry for the simulated plankton groups \citep{baretta1997microbial, geider1997dynamic} and represents the biomass of each functional type in terms of chlorophyll, carbon, nitrogen, and phosphorus, with diatoms also being represented by silicon. 
ERSEM predators consist of three types of zooplankton (mesozooplankton, microzooplankton, and heterotrophic nanoflagellates), with organic material being decomposed by a single type of heterotrophic bacteria \citep{butenschon2016ersem}. 
The model represents three different sizes of detritus (small, medium and large) and three types of dissolved organic matter (DOM: refractory; semi-labile; labile). 
The inorganic component of ERSEM includes nutrients such as nitrate, phosphate, silicate, ammonium, and carbon, as well as dissolved oxygen. The carbonate system is also included in the model \citep{artioli2012carbonate}. ERSEM has been used for many applications including NWES and Mediterranean Sea biogeochemistry reanalyses \citep{ciavatta2016decadal, ciavatta2018assimilation, ciavatta2019ecoregions}, NWES operational forecast \citep{skakala2018assimilation}, and NWES climate projections \citep{wakelin2015modelling, wakelin2020controls, galli2024multi, McEwan2021CMEMS}. 

\begin{table}[h]
    \centering
    \begin{tabular}{l l l} 
        \toprule
        Functional Group & Class/Type & Chemical Components \\
        \midrule
         Phytoplankton  &  Diatoms  & $Chl$, C, N, P, Si     \\
         Functional Types &  Microphytoplankton  & $Chl$, C, N, P  \\
         (PFT)  &  Nanophytoplankton  & $Chl$, C, N, P   \\
           &  Picophytoplankton  & $Chl$, C, N, P   \\
        \hline
        Zooplankton   &  Mesozooplankton  & C   \\
                      &  Microzooplankton  & C, N, P   \\
                      &  Heterotrophic Flagellates  & C, N, P   \\
        \hline
        Bacteria   & -       & C, N, P   \\
        \hline
        Detritus   & Small   & C, N, P   \\
                   & Medium  & C, N, P, Si    \\
                   & Large   & C, N, P, Si    \\
        \hline
        Dissolved Organic Matter    & Labile   & C, N, P   \\
        (DOM)                       & Semi-labile  & C \\
                                    & Refractory   & C   \\
        \hline
        Nutrient       & Nitrate ($NO_3^{-}$)    & N   \\
                        & Phosphate ($PO_4^{3-}$)  & P \\
                        & Ammonium ($NH_4^{+}$)    & N   \\
                        & Silicate ($SiO_4^{4-}$)   & Si \\
        \hline
        Other           & Temperature    &  -  \\
                        & Oxygen $O_2$  & - \\
        \bottomrule
    \end{tabular}
    \caption{Reference table for ERSEM pelagic variables used in this study. Chemical components are represented by the following symbols: $Chl$ is chlorophyll; C is carbon; N is nitrogen; P is phosphorus and Si is silicon. Note that we also use total chlorophyll (denoted as $c$ in this paper), which is a diagnostic variable calculated as the sum of chlorophyll concentrations from all PFT classes. }
    \label{tab:ref_table}
\end{table}

The coupler known as the ``Framework for Aquatic Biogeochemical Models'' (FABM) \citep{bruggeman2014general} allows for the smooth combination of hydrodynamic and biogeochemical models, and is used to couple GOTM with ERSEM in this work. The coupling of GOTM to marine BGC models using FABM has allowed for a wide range of applications that include modelling of phytoplankton growth \citep{kerimoglu2021fabm}, implications of sea-ice BGC for oceanic emissions \citep{hayashida2017implications}, investigations on the highly intermittent spatial variability of phytoplankton on sub-grid scales \citep{mandal20161d} and enhancing stoichiometry in existing BGC models \citep{anugerahanti2021enhancing}.

\subsection{Model configuration and synthetic data setup}
\label{sec:model_config}
\begin{figure}[H]
    \centering
    \includegraphics[width=0.7\textwidth]{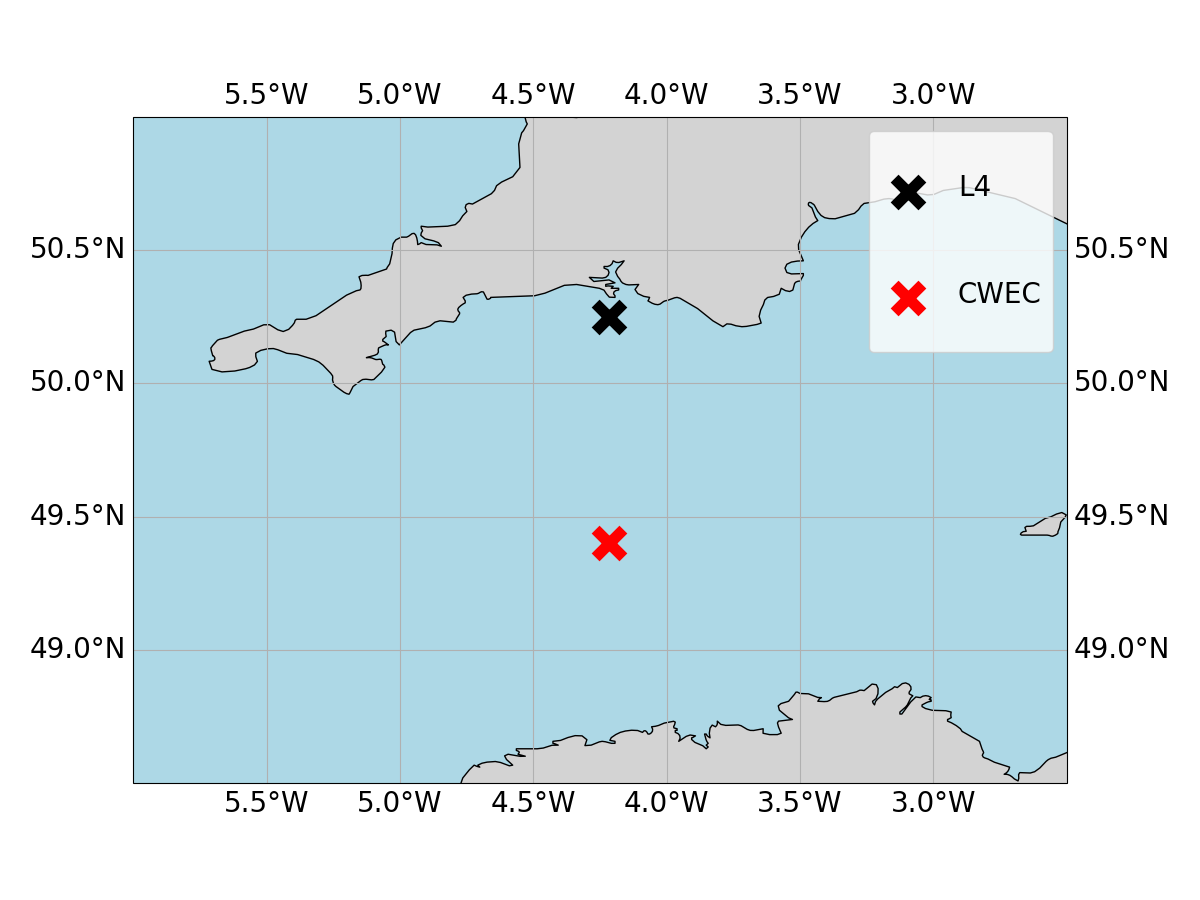}   
    \caption{Map of the Western English Channel, marking the L4 model-training location with a black cross and the CWEC (Central Western English Channel) with a red cross, where we evaluated the model portability.}
    \label{fig:location_map}
\end{figure}

We configure the GOTM-FABM-ERSEM setup for two different locations in the English Channel (see Fig.~\ref{fig:location_map}) and use synthetic observations of each. 
The first location, known as L4 ($50.25\degree\mathrm{N}$, $4.217\degree\mathrm{W}$),  is a highly biologically productive site with seasonally stratified dynamics \citep{pingree1978tidal}, influenced significantly by the outflow of the nearby Tamar and Plym rivers. Nitrate acts as the primary limiting nutrient for phytoplankton growth.
It is monitored by the Western Channel Observatory (WCO) (\href{https://www.westernchannelobservatory.org.uk/}{https://www.westernchannelobservatory.org.uk/}) and SmartSound Plymouth (\href{https://www.smartsoundplymouth.co.uk/}{https://www.smartsoundplymouth.co.uk/}).

Besides the L4 site, we configure a setup for an additional location, that we shall refer to as the Central Western English Channel (CWEC), at $49.40\degree\mathrm{N}$, $4.217\degree\mathrm{W}$. This point is less biologically productive and it is much less influenced by riverine outflow than L4. These differences are evident when looking at the distributions of biogeochemical signals in the models applied at these two locations (see Fig.~\ref{fig:app:behaviour_dist}). The differences make CWEC a reasonable alternative test site for assessing the application of the ML model, and its suitability to generalise the results of this study under different marine BGC conditions.

The physical and biogeochemical models for each location are forced with data appropriate for the study area, using the following datasets:
the General Bathymetric Chart of the Oceans 2023 ($1/240\degree$ resolution) for water depth; the ECMWF ERA5 dataset ($0.25\degree/\mathrm{hourly}$ resolution) for meteorology; the TPXO9-atlas ($1/30\degree$ resolution) for tides; and the World Ocean Atlas 2018 ($0.25\degree$ resolution) for temperature, salinity and nutrient fields for biogeochemical relaxation profiles. 
A nutrient relaxation timescale of $3$ months towards the World Ocean Atlas data is required to prevent significant trends forming that cause the 1D model to gradually accumulate nutrients. This relaxation is significantly longer than the assimilation cycle of $7$ days, and so has little impact on forecast errors. 

Ensemble runs, whether as free runs or for the EnKF (see Sect.~\ref{sec:EnKF}) are configured and run using the Ensemble and Assimilation Tool (EAT) in Python \citep{bruggeman2024eat}. Each ensemble is given a spin-up period of 10 years to settle the biogeochemistry appropriately and provide well-spread initial conditions. 
Each ensemble member uses a signal of temporally correlated random noise to scale the ECMWF ERA wind forcing at the location. The scaling noise signal has a correlation timescale of $7$ days, a mean of $1$ and a standard deviation of $0.3$. The resulting variation in wind strength across the ensemble members increases their spread over time, and prevents ensemble collapse induced by the previously mentioned nutrient relaxation or lack of representation of other error growth processes like horizontal advection, which are absent in a 1D set-up.

\subsection{Data assimilation setups}\label{sec:da_setups}
We examine a total of five data assimilation (DA) setups in this work.  These are conventional DA methods -- namely a simple univariate scheme to reflect how DA is done currently in operational marine BGC systems, and an EnKF for comparison -- to the new schemes that are hybridised with ML techniques.  Before describing each scheme, we give the basic equations of conventional DA, introduce the state vector that the DA uses, the observation type that we will assimilate, and other adjustments that are done post assimilation.

The update equation that is central to DA is sometimes called the best linear unbiased estimator \citep[BLUE, ][]{asch2016data, carrassi2018data} and is given by
\begin{equation}
\label{eq:single_model_update}
\mathbf{x}^{\rm{a}} = \mathbf{x}^{\rm{f}} + \mathbf{K} (\mathbf{y} - \mathcal{H} (\mathbf{x}^{\rm{f}}) ),
\end{equation}
where $\Delta  \mathbf{x}$ is the analysis increment
\begin{equation}
\label{eq:an_inc}
\Delta \mathbf{x} =  \mathbf{K} (\mathbf{y} - \mathcal{H} (\mathbf{x}^{\rm{f}}) ),
\end{equation}
and $\mathbf{K}$ is the Kalman gain matrix
\begin{equation}
\label{eq:KalmanGain}
\mathbf{K} = \mathbf{P}^{\rm{f}} \mathbf{H}^\top (\mathbf{H} \mathbf{P}^{\rm{f}} \mathbf{H}^\top + \mathbf{R})^{-1},
\end{equation}
and where
$\mathbf{x}^{\rm{a}}$ is the analysis (updated) state,
$\mathbf{x}^{\rm{f}}$ is the background state,
$\mathbf{y}$ are the observations,
$\mathcal{H}$ is the observation operator (with Jacobian $\mathbf {H}$),
$\mathbf{P}^{\rm{f}}$ is the background error covariance matrix, and 
$\mathbf{R}$ is the observation error covariance matrix.
The matrix $\mathbf{P}^{\rm{f}}$ is of special interest to this work. Ideally this matrix should be appropriately flow-dependent, but in practice it is often not, such as in many operational schemes.  The purpose of this work is to introduce such flow-dependency to $\mathbf{P}^{\rm{f}}$, or to the analysis increments $\Delta \mathbf{x}$, with ML techniques.

In this work, the state of the system, both $\mathbf{x}^{\rm{f}}$ and $\mathbf{x}^{\rm{a}}$, comprises the surface values of most pelagic variables in ERSEM. We observe only the total chlorophyll, $x_{c}$, at the surface. However, total chlorophyll is a diagnostic variable constructed by summing the chlorophyll concentrations of each phytoplankton function type (PFT, see Table~\ref{tab:ref_table}) that are prognostic variables.

We will only ever have a single surface observation of total chlorophyll at each assimilation step. Our observation operator is therefore always linear ($\mathcal{H}(\bullet)=\mathbf{H}\bullet$) and takes the form:
\begin{equation}
\label{eq:ob_operator}
\mathbf{H} = [1, 0_1, ..., 0_{N}],
\end{equation}
where $N$ is the number of chosen unobserved variables remaining in the state, and so the dimension of the system is $N+1$ with the first element being surface total chlorophyll. Note that this $\mathbf{H}$ has the structure of a row vector as we have only a single observation for each DA update.

The PFTs are excluded from the state, as they are updated after the analysis in Eq.~\eqref{eq:single_model_update}. The PFT chlorophyll updates are computed as a proportion of the background ratio with total chlorophyll:
\begin{equation}    
\label{eq:stoich_chl}
x_{\chi}^{{\rm a}}=x_{\chi}^{{\rm f}}+\frac{x_{\chi}^{{\rm f}}}{x_{c}^{{\rm f}}}\cdot\left(x_{c}^{{\rm a}}-x_{c}^{{\rm f}}\right), 
\end{equation}
where $\chi$ stands for each of the chlorophyll components of the PFTs, and $c$ stands for the total chlorophyll.
Then, the additional chemical components of each PFT, which are carbon, nitrate, phosphate (and in the case of diatoms, silicate as well) are updated according to the background ratios between the chlorophyll of the given PFT and the chemical component as:
\begin{equation}
\label{eq:stoich_chem}
x_{\zeta}^{{\rm a}}=x_{\zeta}^{{\rm f}}+\frac{x_{\zeta}^{{\rm f}}}{x_{\chi}^{{\rm f}}}\cdot\left(x_{\chi}^{{\rm a}}-x_{\chi}^{{\rm f}}\right), 
\end{equation}
where $\zeta$ stands for each of the non-chlorophyll chemical components of a PFT. Which of the remaining elements of the state vector at the surface are updated, and how, depends on the specific DA scheme (see below).

All DA methods here will only directly update the surface layer of the model. However, the rest of the mixed layer is also updated, and uses this same surface analysis increment. Propagating surface analysis increments in this manner makes the reasonable assumption that the behaviour at the model surface is largely representative of behaviour through the mixed layer, and so the resulting analysis increments at the surface are also approximately correct across these additional depth layers. Variables below the mixed layer are not updated.

The specific DA schemes (conventional and ML-based) are now described. A summary of the methods is given in Table \ref{tab:run_types}.

\begin{table}[h!]
    \centering
    \renewcommand{\arraystretch}{1.5} 
    \resizebox{\textwidth}{!}{ 
    \begin{tabular}{>{\centering\arraybackslash}m{1.8cm}>{\centering\arraybackslash}m{4cm}>{\centering\arraybackslash}m{1.6cm}>{\centering\arraybackslash}m{1.6cm}>{\centering\arraybackslash}m{1.6cm}>{\centering\arraybackslash}m{1.6cm}>{\centering\arraybackslash}m{1.6cm}}
    \toprule
    \textbf{Run / scheme} & \textbf{Description / purpose} & \textbf{$c$ variance} & \textbf{$i$ variance} & \textbf{$c$-$i$ correlation source} & \textbf{$\Delta x_{c}$} & \textbf{$\Delta x_{i}$} \tabularnewline
    \midrule
    \multicolumn{7}{c}{\rule{0pt}{5mm} \textbf{Preparation / training runs} \rule{0pt}{5mm}} \tabularnewline
    \hline
    1. Truth run & To synthesise observations and for analysis evaluation & n/a & n/a & n/a & n/a & n/a \tabularnewline
    2.~Ensemble of free-runs & To determine climatological correlations and training for ML-OI & n/a & n/a & n/a & n/a & n/a \tabularnewline
    3. EnKF & Update all chosen surface variables / gold standard run / training for ML-EtE & itself & itself & itself & Eq. (1) & Eq. (1) \tabularnewline
    \hline 
    \multicolumn{7}{c}{\rule{0pt}{5mm} \textbf{Conventional assimilation runs} \rule{0pt}{5mm}} \tabularnewline
    \hline
    4. RUS & Reference univariate scheme (TC + stoichiometrical PFT update) / baseline
    for extensions & climatology & n/a & n/a & Eq. (1) & zero \tabularnewline
    \hline 
    \multicolumn{7}{c}{\rule{0pt}{5mm} \textbf{RUS extension assimilation runs (update to variable $i$ with ML methods)} \rule{0pt}{5mm}} \tabularnewline
    \hline 
    5. ML-OI & ML correlation hybrid & climatology & climatology & ML of free run & As RUS & Eq. (4) with (5) \tabularnewline
    6. ML-EtE & ML end-to-end EnKF emulation & climatology & n/a & n/a & As RUS & ML \tabularnewline
    \hline 
    \multicolumn{7}{c}{\rule{0pt}{5mm} \textbf{RUS extension assimilation runs (update to variable $i$ with non-ML methods)} \rule{0pt}{5mm}} \tabularnewline
    \hline 
    7. CliC & Climatological correlations & climatology & climatology & climatology & As RUS & Eq. (4) with (5) \tabularnewline
    \bottomrule
    \end{tabular}
    }
    \caption{An overview of the different run-types and schemes used in this work. The truth run refers to a single-model run with no updates. We sample synthetic observations from this run and feed these into each DA scheme. The ensemble of free-runs means the model is left to run without assimilation. The EnKF uses an ensemble to model background error covariance in the DA update of all state variables (Sect.~\ref{sec:EnKF}). The RUS is the `univariate' scheme (Sect.~\ref{sec:method_ukmo}), which is used as a benchmark for the performance of other schemes. It updates only the total chlorophyll state variable. The ML-OI predicts background correlations of variables beyond the total chlorophyll with an ANN (Sect.~\ref{sec:method_mlfc}). The ML-EtE predicts the analysis increments of variables beyond the total chlorophyll produced by an EnKF using an ANN (Sect.~\ref{sec:method_mlai}). The CliC is similar to ML-OI but uses purely climatological background statistical estimates of the correlations to update the state of unobserved variables (Sect.~\ref{sec:method_clim}).}
    \label{tab:run_types}
\end{table}

\subsubsection{Reference univariate DA scheme (RUS)}
\label{sec:method_ukmo}
We call our baseline DA method the reference ``univariate'' DA scheme (RUS, Table \ref{tab:run_types}, row 4).  Its purpose is to mimic existing DA systems used by several operational centres \citep{teruzzi2014a3dvarassim, skakala2018assimilation}, although our scheme is not variational. The background error covariances are based on climatological information and so do not adapt to the state.

The RUS is based on an evaluation of Eq.~\eqref{eq:single_model_update}, but only to directly update the total chlorophyll variable.  The simple structure of the observation operator in Eq.~\eqref{eq:ob_operator} means we can rewrite the update Eq.~\eqref{eq:single_model_update} to show how the total chlorophyll (index $c$) is updated from the total chlorophyll observation:
\begin{equation}
    x^{\rm{a}}_c = x^{\rm{f}}_c
    + 
        \frac{
        P_{c,c}^{\rm{f}}
        }
        {P_{c,c}^{\rm{f}} + R} 
        \cdot (y -  x^{\rm{f}}_c),
\label{eq:update_c}
\end{equation}
where $P_{c,c}^{\rm{f}}$ is the background error variance of total chlorophyll. Climatological variances from a long training EnKF run (Sect.~\ref{sec:EnKF}) are used to estimate $P_{c,c}^{\rm{f}}$. Details on the training runs can be found in Sect.~\ref{sec:method_mlfc} and \ref{sec:method_mlai}. Updates to the surface PFT chlorophyll, to the associated chemical components, and throughout the mixed layer are made separately as described at the start of Sect.~\ref{sec:da_setups}.

Note that we call this scheme ``univariate'' as only a single variable (total chlorophyll) is updated according to the background and observational errors as described in Eq.~\eqref{eq:update_c}.  All further DA schemes described in this work (apart from the EnKF below, which uses ensemble-derived covariances) start with an update of the total chlorophyll using Eq.~\eqref{eq:update_c}, and will attempt to update additional pelagic variables using the new ML-based approaches.

\subsubsection{The EnKF-based scheme}\label{sec:EnKF}
The stochastic EnKF scheme, see e.g., \cite{evensen2003ensemble}, approximates the update Eqs.~\eqref{eq:single_model_update} and \eqref{eq:KalmanGain} with an ensemble to estimate the flow-dependent background error covariance matrix $\mathbf{P}^{\rm{f}}$, Table~\ref{tab:run_types}, row 3. For each ensemble member there is a different update and a different perturbed observation (the perturbations are sampled from the normal distribution $\mathcal{N}(0,R)$). The EnKF updates all elements of the surface state described previously using the ensemble version of Eq.~\eqref{eq:single_model_update}, but still performs the stoichiometric balancing scheme and duplication of the analysis increments from the surface throughout the mixed layer, as described in Sect.~\ref{sec:da_setups}.

The implementation of an EnKF to this problem is relatively expensive, but provides a ``gold standard'' comparison DA method, plus it provides valuable training data for the ML model, as specified in Sect.~\ref{sec:method_mlai}.  The accuracy of the EnKF will depend on the number of ensemble members. This is discussed in Sect.~\ref{sec:nitrate_only}.

\section{Hybrid machine learning data assimilation for marine biogeochemistry}
\label{sec:new_schemes}
In this section, we describe how we hybridise the DA, described above, with ML to provide flow-dependent estimates of the statistics/increments that are better than the climatological values. In particular, we take two approaches that differently replace parts of, or fully, the update equation. We now show the mathematical framework that the ML schemes will emulate, which is derived from Eqs.~\eqref{eq:single_model_update}-\eqref{eq:ob_operator}.

\subsection{Mathematical framework for the ML-based DA schemes}
\label{sec:framework_ML}
The ML-based DA schemes are summarised in Table~\ref{tab:run_types}, rows 5 and 6. They both build upon RUS, extending it to become multivariate. The total chlorophyll analysis is computed using the RUS update Eq.~\eqref{eq:update_c}, while the remaining variables (potentially $1\leq i\leq N$) have updates according to
\begin{equation}
    x^{\rm{a}}_i = x^{\rm{f}}_i 
    + 
    \underbrace{
        \frac{
        P_{i, c}^{\rm{f}}
        }
        {P_{c, c}^{\rm{f}} + R} 
        \cdot (y -  x^{\rm{f}}_c) ,
    }_{\textrm{analysis increment}}
    \label{eq:simplified_enkf}
\end{equation}
where $P_{i,c}^{\rm f}$ is the background error covariance between variable $i$ and total chlorophyll defined as 
\begin{equation}
\label{eq:covariance_calc}
    P^{\rm{f}}_{i,c} = {COR}_{i,c} \cdot \sigma_i \cdot \sigma_c ,
\end{equation}
where ${COR}_{i,c}$ is their forecast error correlation, and $\sigma_i$ and $\sigma_c$ are their respective background error standard deviations.
In Eq.~\eqref{eq:simplified_enkf} the analysis increment of the update is labelled.  

An important aspect of any DA scheme is its ability to adapt with the flow. A conventional way to introduce flow-dependency is via Monte Carlo-like methods such the EnKF, which comes with substantial computational cost. 
The two proposed ML-DA schemes below are designed with the above in mind and provide flow-dependency cost-effectively without the need for an ensemble (apart from at the training stage, as shall be clarified). The two ML-DA schemes are described in Sections~\ref{sec:method_mlfc} and \ref{sec:method_mlai}.  The specific details of their architecture are given in Sect.~\ref{sec:ML_architecture}.

\subsection{Hybrid machine-learning optimal interpolation (ML-OI)}
\label{sec:method_mlfc}
This approach first updates the observed total chlorophyll and associated PFTs in an identical manner to the RUS described in Sect.~\ref{sec:method_ukmo}. The first ML method uses the scaled background state (see below) to predict the state-dependent correlations between observed and unobserved quantities in Eq.~\eqref{eq:covariance_calc}. Together with climatologically estimated values of $\sigma_i$ and $\sigma_c$, the correlations are substituted into Eqs.~\eqref{eq:covariance_calc} and then \eqref{eq:simplified_enkf} to provide updates to the unobserved variables in the system. We call this approach ML-OI (``optimal interpolation'', Table~\ref{tab:run_types}, row 5). The background state of each variable that is input into the ML-OI model to predict ${COR}_{i,c}$ is scaled according its climatological maximum for the given location (e.g., L4 in Fig.~\ref{fig:location_map}), assuming that the correlative relationship between variables will be similar, even if the scale of seasonal variability is different. Standard deviations are each a statistic of a single variable, which we assume is easier to capture climatologically than a correlation is. The resulting surface increments of the unobserved variables  are then propagated to the other levels in the mixed layer as described previously.

While the field of hybrid ML-DA is growing rapidly, there exists relatively few works in which ML-predicted background error covariances are so closely coupled with existing DA systems. However, a few particularly relevant examples stand out such as \cite{ouala2018neural}, in which a Kalman-like analysis update is applied to satellite-derived sea surface temperature fields using artificial neural network (ANN)-predicted background error covariances. Additional examples of this can be seen in \cite{sacco2022evaluation}, which aim to learn different sources of uncertainty using ANNs on both toy models and sea level pressure forecasts. Further work  \citep{sacco2024line} uses an EnKF to generate flow dependent background error covariances, and then learns them using a convolutional neural network.

In order to generate training data for this approach, we run a 100-member ensemble of free-runs, configured according to Sect.~\ref{sec:model_config} (Table~\ref{tab:run_types}, row 2). 
We generate training samples at seven day intervals across these free-runs, covering the period from 2000-2014. The features are the surface states of individual ensemble members at a given time, across all pelagic model variables. 
For the first application of ML-OI in Sect.~\ref{sec:nitrate_only}, the targets are time dependent/ensemble-derived correlations between total chlorophyll and nitrate. In the later application in Sect.~\ref{sec:result_extended_variable_set} onwards this is extended from just nitrate to a wider set of variables.

\subsection{End-to-end machine learning of EnKF updates (ML-EtE)}
\label{sec:method_mlai}
This approach again first updates the observable total chlorophyll and associated PFTs in an identical manner to the RUS described in Sect.~\ref{sec:method_ukmo}. Nevertheless, as opposed to ML-OI, ML is used here to predict the analysis increments for unobserved variables, given the analysis increment of the observed variable (total chlorophyll) and the complete background state. This obviously requires running a DA system to learn from. This is achieved here using the updates produced by an EnKF training run (see below). We call this approach ML-EtE  (``end-to-end'', Table~\ref{tab:run_types}, row 6) emulation of an existing DA system. 

In ML-EtE, we assume that the essential properties of each statistical object that creates an analysis increment, such as covariances and observation uncertainty, can be more effectively captured by directly predicting the analysis increment rather than predicting every component individually and allowing errors to compound across multiple independent predictions that are then combined into a single value. The resulting surface increments of the unobserved variables  are then propagated to the other levels in the mixed layer as described previously.

This approach follows other work in a similar vein, such as work by \cite{bonavita2020machine}, which uses ANNs to emulate the main features of an operational weak-constraint 4D-Var scheme. Let us reiterate however that ML-EtE replaces, in an end-to-end fashion, only the analysis step of the EnKF, and not the full forecast-analysis cycle (including therefore the dynamical model). As mentioned in the introduction, a similar scope characterises the work by \cite{bocquet2024accurate}.

To generate the training data for this approach, we first generate a nature run for the training period (Table~\ref{tab:run_types}, row 1), to generate synthetic surface observations of total chlorophyll concentration at weekly intervals. The observation uncertainty is equal to 10\% of the observed value. These are then assimilated into the EnKF run over the same period. The features of each training sample consist of an individual ensemble member's background state and its corresponding total chlorophyll increment from the EnKF run. The targets are the corresponding analysis increments for the unobserved variables.

As mentioned, emulating analysis increments of existing DA systems is shown in existing non-marine BGC works, in particular in relation to estimating and correcting model error \citep{bonavita2020machine,brajard2020combining, gregory2024machine}. In all cases, it relies on having an existing DA system in place, or, as in the above mentioned works, on a robust reanalysis. 
The availability of a reanalysis is a clear obstacle of this approach. We shall discuss this further in our conclusion. Here we are primarily interested in studying its feasibility and ability to learn successfully the EnKF updates.

\subsection{Machine learning architecture}
\label{sec:ML_architecture}
Each ML model is a fully connected ANN optimized using AutoKeras \citep{jin2019auto} over 100 trial configurations. AutoKeras uses Bayesian optimisation in a network search algorithm to determine optimal hyperparameters such as layer depth, layer width, dropout rate, learning rate, and optimiser selection. The input features of each model are standardised to have unit variance and a mean of zero. 

Each ML approach is tested in the following scenarios: (1) a set-up where we choose to update only nitrate, (2a) a set-up where we update the full set of pelagic variables, and (2b) a set-up where we update a partial set of the pelagic variables, eliminating poorly estimated variables based on the results of (2a).

\subsection{Purely climatological updates}
\label{sec:method_clim}
A further non-ML-based scheme is used to update the extended range of variables to mirror ML-OI, but using only climatological correlations derived from the EnKF run (CliC, Table \ref{tab:run_types}, row 7) over the training period. This serves as another comparison point, a benchmark, to check whether the additional complexity of an ML model is needed.

\subsection{Skill metric and machine learning model evaluation}

\subsubsection{Skill metric}
\label{sec:skill_metrics}
For a system that runs for $\tau$ cycles, we represent the trajectory for a member $i$ of ensemble $X$ at cycle $t$ as $X^i_t$.
The truth is denoted as $T_t$.
The expected RMSE (root mean square error) over $M$ ensemble members (or a set of $M$ single-model runs), is calculated as:
\begin{equation}
    RMSE = \frac{1}{M}\sum_{i=1}^M \sqrt{ \frac{1}{\tau}\sum_{t=1}^\tau  (X^{i}_t - T_t)^2}.
    \label{eq:expected_error}
\end{equation}
This is a sensible metric to use when calculating the expected error across a set of independent single-model runs, such as in the RUS, CliC, ML-EtE and ML-OI approaches. It is also convenient for calculating the expected error of the ensemble members in the EnKF runs.

\subsubsection{SHAP analysis}
\label{sec:SHAP_metric}
Shapley values are a well known and widely used metric for understanding the importance and contributions of individual input features in ML models \citep{lundberg2017unified}. A Shapley value represents the average marginal contribution of a feature across all possible subsets of features, ensuring a fair allocation of importance.
In this work, we use Kernel SHAP (SHapley Additive exPlanations) as a model-agnostic approach to estimate mean Shapely values across a dataset. Kernel SHAP approximates Shapley values by training a weighted linear model on perturbations of the input data. 
By calculating the mean absolute Shapley values, we measure the magnitude of influence for individual features to the model's predictions.

By understanding the importance of each input feature, we gain insight into the correlative links of dynamical behaviour in the system. This can help to identify how-and-when a model will translate well to new conditions. For example, if the primary predictive feature of an ML model is similar in two separate locations, one trained and one unseen, then we may expect the ML model to perform reasonably well in the new scenario, even if the other non-predictive features exhibit an entirely different distribution.
We emphasise that we cannot infer causality from this analysis alone but understanding the data-driven feature importance and feature contribution for an ML model, combined with expert understanding of the system dynamics, can help to unveil connections and insights into the complex processes of the marine BGC model.

It also also worth noting that these metrics can also be used for feature selection with the idea that if a feature contributes little-to-nothing to the predictions, it can probably be eliminated from the feature set. This then requires the expensive processes of iteratively re-training and re-testing the neural networks and so is not an avenue that we explore in this work.
SHAP is somewhat limited in the presence of highly correlated features because Shapley values assume independent feature contributions. This can lead to arbitrary or shared attributions when features provide redundant information, making it difficult to disentangle their true individual impacts. However, the correlation structures of the MGBC have been studied previously \citep{higgs2024investigating}, and so can be more effectively accounted for during analysis.

\section{Results and discussion}
\label{sec:discussion}
\subsection{System dynamics}
\label{sec:sys_dynamics}
As discussed in Sect.~\ref{sec:model_config}, the L4 location is a highly biologically productive site with seasonally stratified dynamics. Nitrogen is a key component of organic matter and is generally the limiting nutrient to primary production by phytoplankton in coastal marine ecosystems \citep{national2000clean}. This leads to a strong, exploitable dynamical link between phytoplankton and nitrate that varies with a clear seasonal cycle. This cycle can be broken down into three distinct regimes across any given year:
\begin{itemize}
    \item The {\it bloom regime} can occur throughout spring (from March until May), and is the period when phytoplankton reaches its yearly maximum. During this time, light levels no longer limit phytoplankton growth and there is a high availability of nutrients that have accumulated in the water column during the ``light-limited'' period.
    \item The {\it nutrient-limited regime} refers to the period roughly spanning from early summer until late September where nutrients, and more specifically nitrate, have been exhausted by the phytoplankton during bloom and so concentrations are generally very low. During this time, phytoplankton relies on processes such as storms to mix nutrients into the upper water column. Consequently, phytoplankton growth is sporadic and less intense than during the spring bloom.
    \item The {\it light-limited regime} typically describes a fully mixed water column, which approximately spans the period from October to the start of the next spring bloom. Here, there is little-to-no phytoplankton growth due to the reduced light-levels, meaning nutrients are mixed throughout the water column without being used by the phytoplankton. During this period, phytoplankton concentrations are very low and mostly decoupled from nutrient dynamics.
\end{itemize}

\begin{figure}[H]
    \centering
    \includegraphics[width=0.99\textwidth]{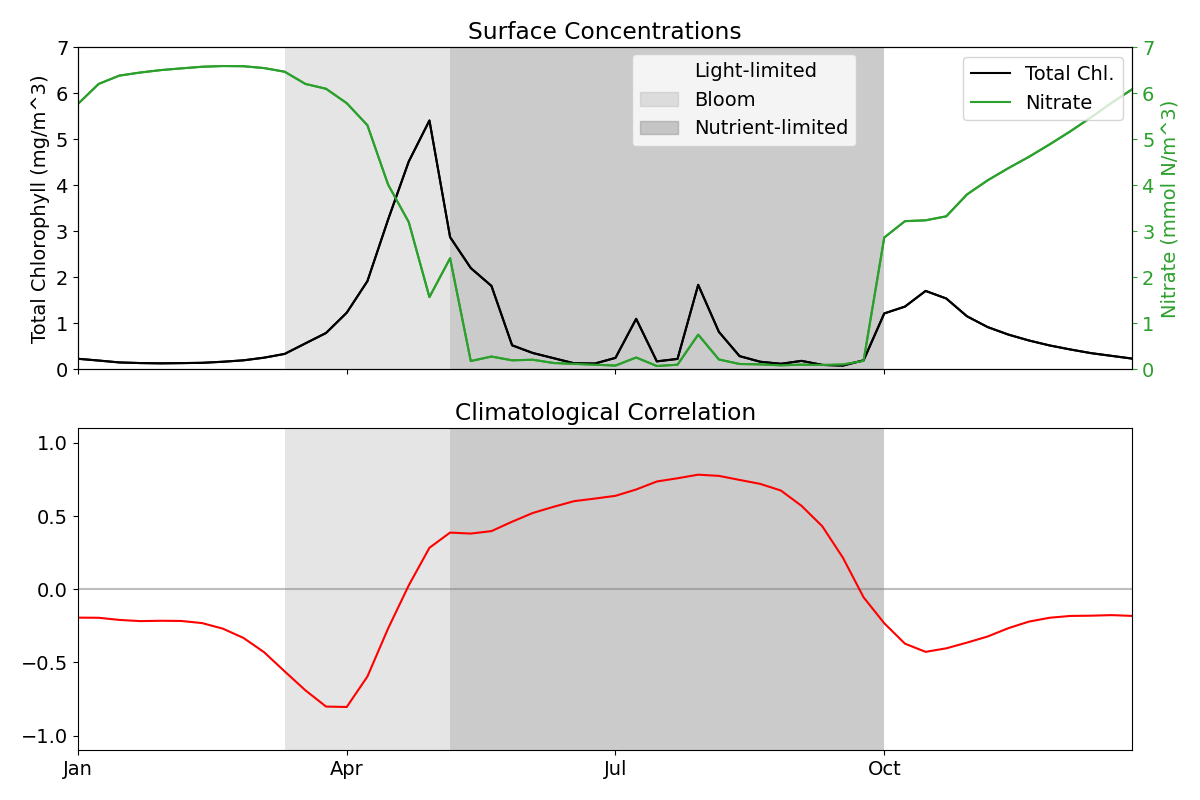}   
    \caption{The top panel shows a time series of surface concentrations of total chlorophyll (black) and nitrate (green) for an arbitrary year at the L4 location. The bottom panel shows the climatological correlation between total chlorophyll and nitrate, calculated across the 2000-2014 training period.
    Shading indicates the dominant seasonal system regimes: ``light-limited'' (white), ``bloom'' (light grey) and ``nutrient-limited'' (dark grey).
    }
    \label{fig:regime_highlight}
\end{figure}

\subsection{Prediction and update to a single pelagic variable}
\label{sec:nitrate_only}
\begin{figure}[H]
    \centering
    \includegraphics[width=0.99\textwidth]{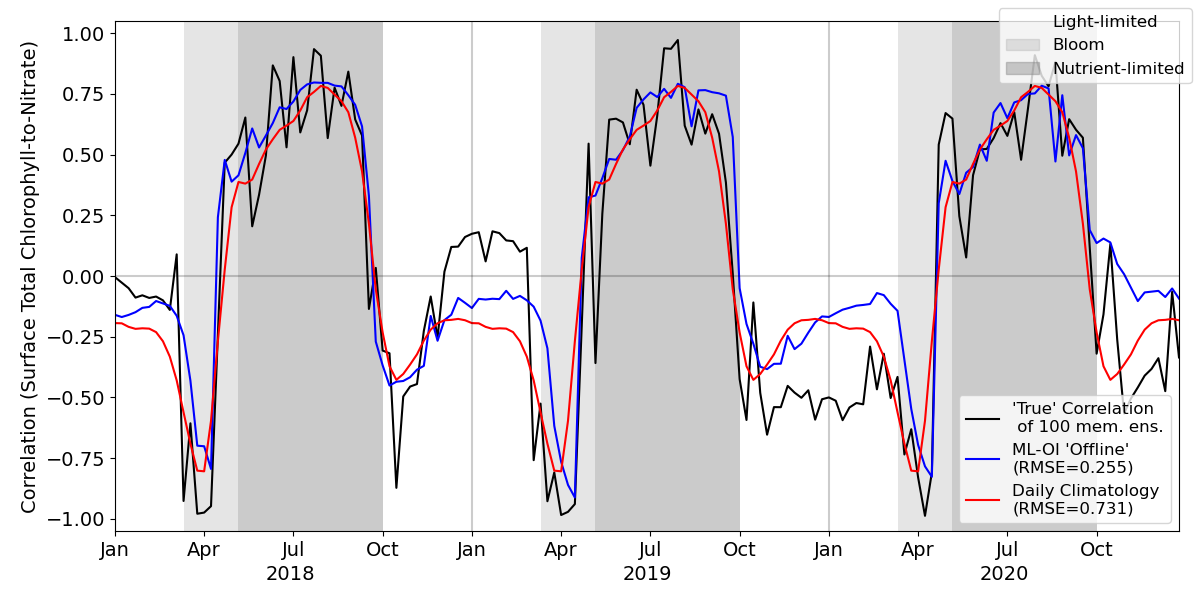}   
    \caption{Predictions for correlation between total chlorophyll and nitrate, at weekly intervals across the 3-year offline test period for the ML-OI approach. The ``true'' correlation (black) is calculated from the 100-member free-run ensemble (Table~\ref{tab:run_types}, row 2). The predictions by ML-OI (blue) are shown, with the RMS difference to the true correlation of 0.255. A daily climatology of correlations has also been calculated from the training data (red), with an RMS difference of 0.731. The seasonal regimes of Figure~\ref{fig:regime_highlight} are repeated.}
    \label{fig:offline_test}
\end{figure}

In this section, we explore the performance of ML-OI and ML-EtE in updating only nitrate as an unobserved variable. Recall however that the observed total chlorophyll and associated PFTs are updated according to the RUS scheme in Sect.~\ref{sec:method_ukmo}. 
We choose nitrate for these initial experiments because it is a limiting nutrient at the L4 location (see Fig.A2 in the Appendix) , and therefore has a clear, explainable relationship with total chlorophyll as discussed in Sect.~\ref{sec:sys_dynamics} (see also Fig.~\ref{fig:regime_highlight}). Since nitrate is the key driver limiting primary production among nutrients, addressing it through DA can have a significant knock-on effect on the whole model state. Thus this also provides us a more understandable proof-of-concept with reduced complexity to analyse initially, before we later extend the updates to more than $30$ additional pelagic variables (see Table~\ref{tab:ref_table}) in a higher complexity scenario.
Figure~\ref{fig:offline_test} shows the correlation between total chlorophyll and nitrate as a function of time in the period 2018-2020 for the ``offline'' ML-OI experiment. Offline refers here to a setup in which the ML-OI analysis is not then used as initial condition for the next forecast, and so it does not impact successive DA cycles. The performance of ML-OI is compared to  the ``true correlation'' computed over an ensemble of 100 members and to the correlation estimated using daily climatology.

The input features (i.e. model background states) and target correlations for ML-OI are taken from the 100-member free-run. As it will become clear below, this is a sufficiently large ensemble to provide accurate correlation estimates. During this test period, the ML model ($RMSE=0.255$) vastly outperforms the climatological statistics estimates ($RMSE=0.731$), which is an approximate $60\%$ improvement. 

The ML-OI model makes clear improvements over the climatological estimates at several points in the yearly cycle. Firstly, it better estimates the highly distinctive correlative pattern between total chlorophyll and nitrate during the bloom regime. This pattern consists of a sharp drop to a strongly negative correlation, before an almost instantaneous increase to a strong positive correlation. 
These correlation patterns can be simply explained. During the bloom, phytoplankton growth exhausts nutrients, leading to negative correlations between chlorophyll and nutrients, whilst the end of the bloom, and the following period, phytoplankton growth is nutrient limited, leading to positive correlation.
The precise timing of the bloom (and hence this correlation pattern) has notable inter-annual variability -- varying within a period of approximately 5 weeks each year, in this model. 
The climatological correlations estimate this pattern poorly as they are smoothed over this period of inter-annual variability, but the ML-OI model, which predicts correlations from the state of the marine BGC model, captures the pattern much more accurately.  
Also, during the nutrient-limited regime, we see a generally strong positive correlation between total chlorophyll and nitrate, which has some local variability primarily driven by changes in wind strength, such as a weather front passing over the location and mixing nutrients into the surface. The ML scheme captures some of the local variability in this ``true'' signal and so responds more accurately these changes in state.
We also see that during the light-limited regime, the system can exist in either a ``weakly positive or no correlation'' state, or a ``moderately negative correlation'' state. During this time, total chlorophyll and nitrate are generally decoupled, and the ensemble concentrations in total chlorophyll are very near zero, meaning the data assimilation has very little impact at this time of year. The latter point means that better correlation predictions at this time of year are less likely to result in any great improvement to the system, as there is weak relationship between chlorophyll and nitrate DA increments. However, updates at the start of this period could be important, as the resulting store of nitrate in the upper water column could have dynamical impact in later DA cycles when light is no longer limiting and the next bloom period starts. 

\begin{figure}[H]
    \centering
    \includegraphics[width=0.99\textwidth]{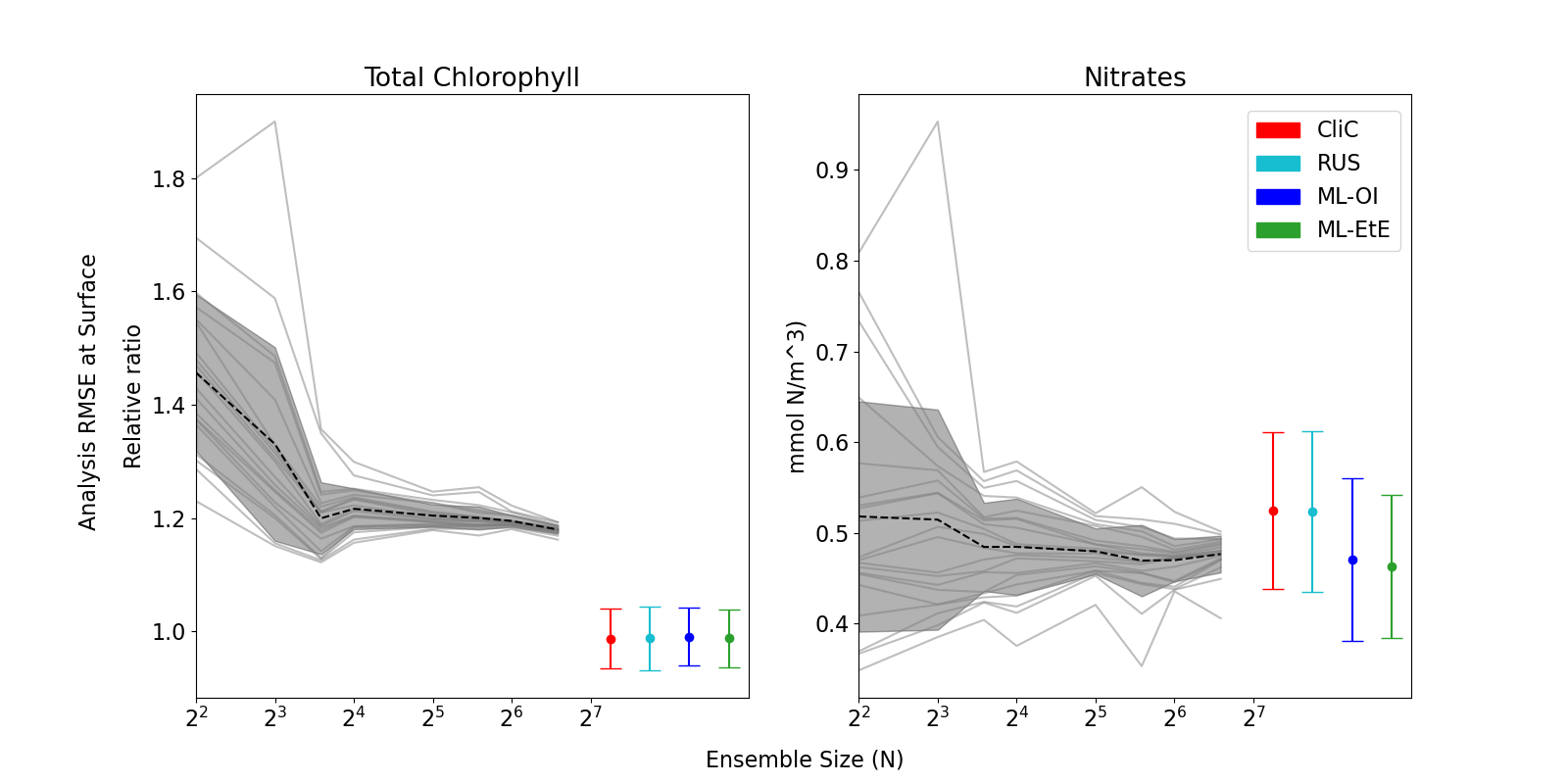}   
    \caption{The relationship between analysis RMSE (Eq.~\eqref{eq:expected_error}) and ensemble size for EnKFs with different ensemble sizes, as well as the performance of the different single-model run schemes. 
    The left panel shows the RMSE of the observed variable, total chlorophyll, normalised relative to the observational error. The right panel shows the RMSE of the unobserved variable, nitrate. 
    The black dashed line represents the mean expected ensemble member error from 20 repeat experiments of an EnKF at increasing ensemble sizes, with the shaded grey area indicating $\pm 1$ standard deviation. 
    The mean error and $\pm 1$ standard deviation of 64 independent single-model runs are also given for each of the methods summarised in Sects.~\ref{sec:da_setups} and \ref{sec:new_schemes}.
    }
    \label{fig:analysis_errors}
\end{figure}

After demonstrating the capability of the ML model to predict chlorophyll-nitrate relationship in an ``offline'' setting in Fig.~\ref{fig:offline_test}, in Fig.~\ref{fig:analysis_errors} we compare the performance of a standard EnKF at different ensemble sizes with the schemes previously summarised in Sect.~\ref{sec:da_setups} and \ref{sec:new_schemes}. This is done in an ``online'' setting, so that any update to the system can have dynamical impact on later DA cycles as the model integrates forward in time.

Figure~\ref{fig:analysis_errors} displays the performance of the EnKF, the RUS scheme, the climatological statistics scheme (CliC), and the ML schemes. Each panel shows the mean expected error of ensemble members for ensemble sizes ranging from $4$ to $96$.
In the left panel for total chlorophyll, the relative RMSE is calculated as a percentage of the observation error. 
The EnKF achieves a near-optimal performance at an ensemble size $>16$, after the mean expected error of ensemble members reaches a plateau with increasing size. The relative analysis error of total chlorophyll is normalised according to observational error. 
exceeds a value of $1$ as we are measuring expected error of ensemble members, not error to the ensemble mean, as described in Sect.~\ref{sec:skill_metrics}. We also see, in the right panel, that the error decreases with ensemble size for the unobserved nitrate.

As expected, the analysis error in the observed total chlorophyll is generally comparable across each scheme because they all use the same method, the RUS scheme of Sect.~\ref{sec:method_ukmo}, to update the observed total chlorophyll. 
However, there are more noticeable differences in the schemes that extend the updates to nitrate as well. In this, we can clearly see that both the RUS scheme (no update to nitrate) and the CliC scheme  (update of nitrate using climatological covariances in Eq.~\eqref{eq:covariance_calc}) perform similarly poorly in improving analysis error of nitrate -- meaning that the information provided by the observation has not propagated well to the unobserved variable. In contrast to this, both ML approaches result in a significant improvement in performance, reducing analysis error by between $8-12\%$. This means that the information from observations can effectively propagate to the unobserved variables in a single-model run, without the need for an expensive ensemble to model the statistics at run time. Also, this indicates that the improvement in correlation prediction shown in the offline experiment of Fig.~\ref{fig:offline_test} translates into the entirely online testing period, where the updates of a given DA cycle will feed into subsequent cycles.

\begin{figure}[H]
    \centering
    \includegraphics[width=0.99\textwidth]{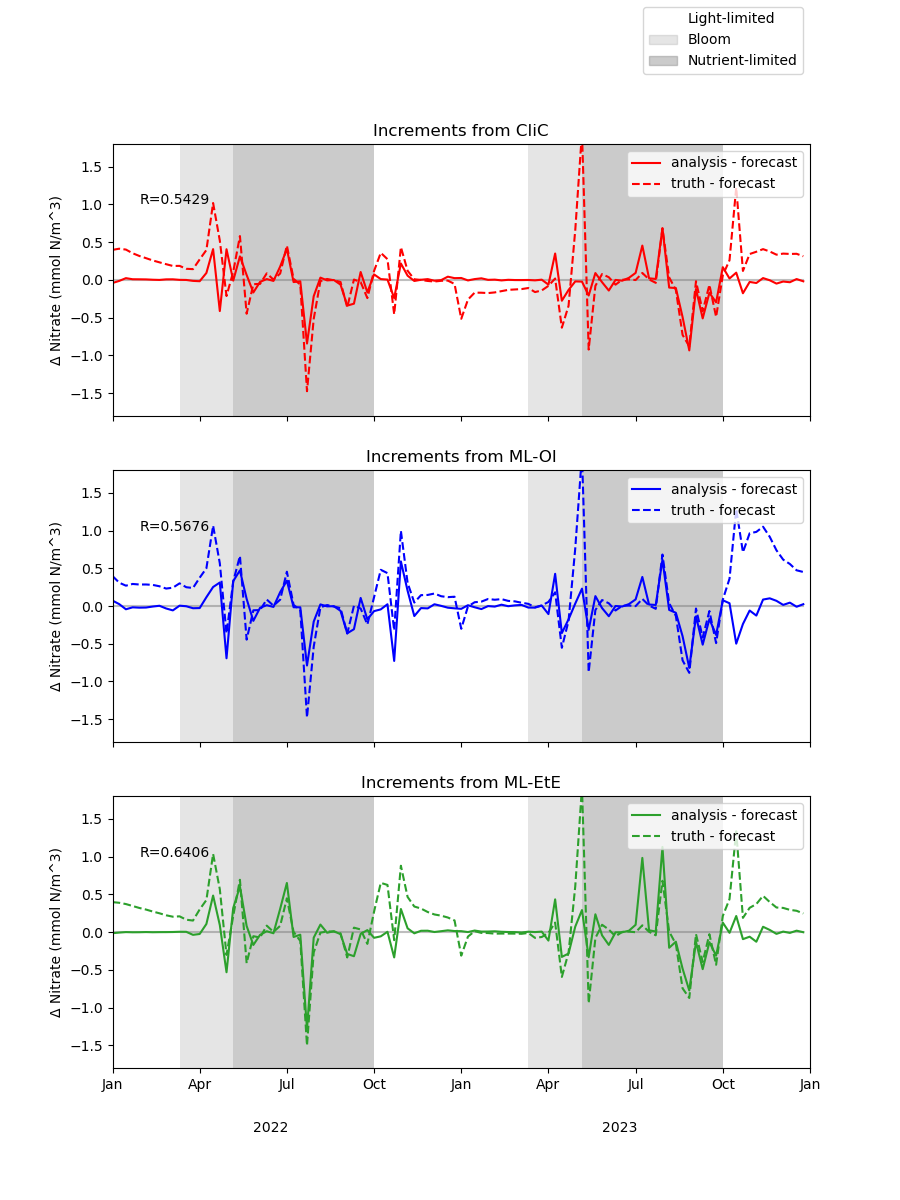}   
    \caption{ A comparison of nitrate analysis increments produced in a single-model run in ``online'' cycled-DA.
    The first panel shows the analysis increments made during the climatological correlations CliC run (solid red) and the difference between the background state and the truth (dashed red).
    The second panel shows the analysis increments made by the ML-predicted correlations ML-OI run (solid blue) and the difference between the background state and the truth (dashed blue).
    The third panel shows the analysis increments predicted directly by the ML-EtE run (solid green) and the difference between the background state and the truth (dashed green).
    In each panel the $R$-value represents the correlation between the analysis increments and the difference between the background and the truth.
    Shading indicates the system regimes previously outlined in Sect.~\ref{sec:sys_dynamics}: ``light-limited'' (white), ``bloom'' (light grey) and ``nutrient-limited'' (dark grey).
    }
    \label{fig:increment_detail}
\end{figure}

These single-model schemes are then investigated further in Fig.~\ref{fig:increment_detail}, looking at the analysis increments generated in the ``online'' setting, and their differences to the truth.

While Fig.~\ref{fig:analysis_errors} shows that they improve on average, Fig.~\ref{fig:increment_detail} gives some details on when improvements are made. The runs shown here receive the same observations of the truth, and use the same initial conditions and forcing. However, the cycled ``online'' DA implies that the background state of a given time step will differ between methods. Nevertheless, we can see when ML-OI, or ML-EtE, make improvements over CliC. A clear example of this is the improved predictions during the bloom period, where the ML-predicted correlations provide a series of increments that are much closer to the truth than the climatological increments. This shows that both ML methods are able to react to the timing of the bloom event much more accurately than climatology can. During the nitrate-limited period, we generally see comparable performance across the increments, as the expected correlations are generally high. However, the ML-EtE approach seems to more accurately capture the largest analysis increments. 
We can also see that each approach makes little to no adjustment during the light-limited regime. The increments in ML-OI and CliC seem to introduce some noise around this time period, while the ML-EtE seems to more reasonably predict negligible increments. While these increments are unlikely to have any major impact on the system, it is interesting to note that the increments from ML-EtE approach, which replicates the end-to-end process of the EnKF increments, seem to more accurately reflect the expected ``decoupling'' of total chlorophyll to nitrate at this time of year. This suggests that ML-EtE has successfully learnt the analysis increments.

\begin{figure}[H]
    \centering
    \includegraphics[width=0.99\textwidth]{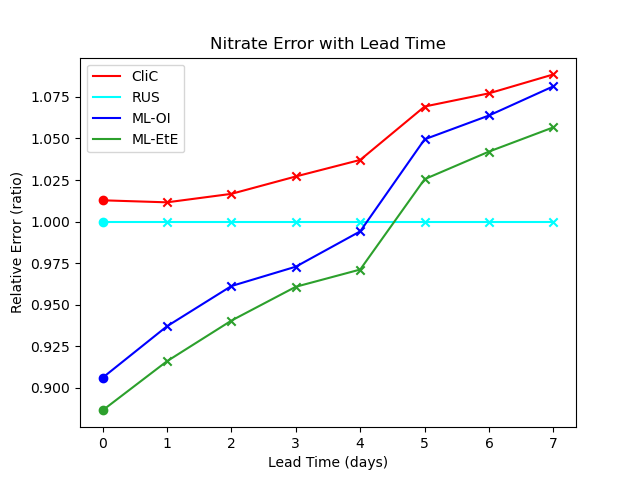}   
    \caption{The forecast RMSE of each scheme relative to that of the RUS scheme at daily lead time increments. For each scheme, the dot indicates the relative analysis error, while the crosses shows the relative forecast error for each day of lead time until a maximum lead time of 7 days - which is the total time between observations of total chlorophyll.
    }
    \label{fig:lead_time}
\end{figure}

Finally, Fig.~\ref{fig:lead_time} shows analysis and forecast errors in nitrate in each scheme where errors are normalised against the error in the RUS scheme. 
The daily climatological correlations (CliC, red line) degrade the analysis error and then make worse forecasts at every lead time when compared to the RUS scheme, which does not update the nitrates at all.
As previously noted, both ML approaches provide an analysis state that is approximately 8-12\% better than the not-updated RUS nitrate. Improved forecasts then persist for approximately 4-5 days of lead time, only reaching an increased relative error after 5 days. For all lead times, ML-EtE outperforms ML-OI. While this is a net benefit to the forecasts of the system, it highlights the difficulty with partially updating a highly non-linear system. In this, it is clear that each attempt to update the nitrate results in an eventual error growth beyond simply not updating the system. 
Part of this could stem from the role of nitrate as a limiting nutrient; in that it is either available to allow phytoplankton growth, or not. This means that when predicting an increment for nitrate, we can know that some nitrate should be present or not, but a precise, continuous quantity that should be added or removed is not information that can necessarily be inferred from the observation of total chlorophyll. However, this error growth could also result from the analysis increments introducing some additional imbalance in other quantities of the system that also need correcting, and the complex marine BGC processes are inter-dependent. These imbalances and forecast error growths are discussed further in the following Sect.~\ref{sec:result_extended_variable_set}, when updating the additional marine BGC variables.

In the context of operational systems, such as those implemented by the UK Met Office, total chlorophyll is assimilated on a daily cycle, and then a forecast is produced for up to six days of lead time from these improved initial conditions. These results imply that there are huge gains to be made not only in short term forecasting (before errors saturate again), but also in reanalysis products that assimilate data with higher frequency, as the ML approaches substantially outperform the RUS scheme at this point.

\subsection{Extending the set of updated variables}
\label{sec:result_extended_variable_set}
\begin{figure}[H]
    \centering
    \includegraphics[width=0.99\textwidth]{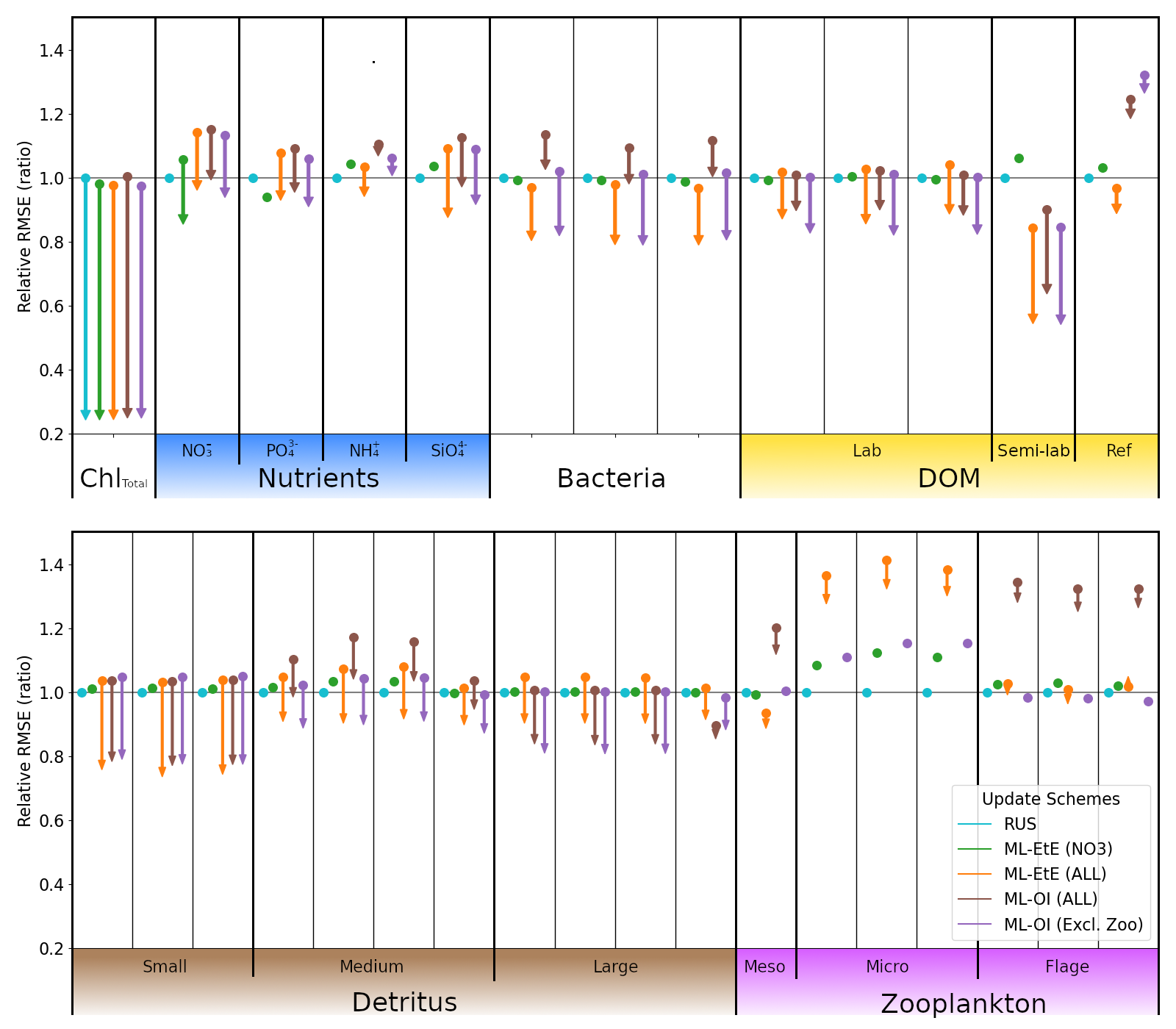}   
    \caption{A comparison of the different schemes implemented to update the various components of the ERSEM BGC model at the L4 location. 
    Forecast RMSEs are shown with dots, and the corresponding arrows indicate the analysis RMSEs (no arrow indicates the variable is not updated by the scheme).  
    All RMSEs are relative to the RMSE in the RUS scheme shown in cyan, which only updates the total chlorophyll and PFTs as described in Sect.~\ref{sec:method_ukmo}. 
    The RMSEs of the ML-EtE (NO3) scheme, green, are from the same experiment shown previously in Sect.~\ref{sec:nitrate_only}, and is used as another comparison point for the extended schemes.
    The ML-EtE (ALL) and ML-OI (ALL), orange and brown respectively, extend the ML schemes described in Sects.~\ref{sec:method_mlai} and \ref{sec:method_mlfc} to update all other pelagic variables. 
    Finally, ML-OI (Excl. Zoo) (purple) updates all pelagic variables, excluding the zooplankton types.
    The chemical components of each variable class/type follow the same order (left to right) as Table~\ref{tab:ref_table}. 
    }
    \label{fig:full_variable_extension}
\end{figure}

In this section, we demonstrate the additional benefit of predicting updates not just for nitrate, but for nearly all marine BGC variables. 
In Fig.~\ref{fig:full_variable_extension}, we compare the different ML approaches for updating an extended set of unobserved marine BGC variables, as well as the previous system that only updates nitrate.

The RUS scheme, described in Sect.~\ref{sec:method_ukmo}, is used as a benchmark for the extended schemes, and so values shown in Fig.~\ref{fig:full_variable_extension} are RMSEs for 7-day forecasts relative to the RMSE of the RUS method. Again, we recall that the RUS does not update any variables beyond total chlorophyll (shown) and its constituent PFTs (not shown).
The ML-EtE (NO3) scheme (green), which updates only nitrate, is carried over from the previous section (as it performed best), to act as another point of comparison for the extended schemes. Before discussing the extended schemes, we can see from Fig.~\ref{fig:full_variable_extension} the dynamical impact that the updates of ML-EtE (NO3) have on other (i.e. non-updated) marine BGC variables in the system. Generally, the change in RMSE for these non-updated variables is very small, with the largest improvement being to phosphate and the largest degradation to zooplankton types -- particularly microzooplankton. While this shows that updating a key nutrient, such as nitrate, can have wider impact on the system through dynamical adjustment, the generally beneficial results of the extended schemes (discussed below) point towards needing a DA system that can make reasonable adjustments to a wider set of marine BGC variables.

Our next scheme, ML-EtE (ALL) (orange), again follows the approach described in Sect.~\ref{sec:method_mlai}, but extends updates to all shown pelagic variables by predicting analysis increments directly from each background state and total chlorophyll increment. 
In this L4 setup, this is generally the best performing scheme, improving unobserved forecast and analysis RMSEs by between $10-50\%$. The most notable exceptions are the zooplankton which, despite having analysis increments in the correct direction, still return noticeably higher forecast/analysis RMSEs than most other schemes. Zooplankton have more interactions with other system components, existing at a higher trophic level, which result in a wider range of uncertainty for their behaviour. This also suggests they have generally weaker correlations with total chlorophyll.

The ML-OI (ALL) scheme (brown) described in \ref{sec:method_mlfc}, extends updates to all shown pelagic variables by predicting the inter-variable correlation from the background state only. This prediction is then combined with daily varying climatological variances to update the marine BGC state. This method is also shown to be effective, generally providing similar behaviour to the ML-EtE (ALL) scheme. It also suffers from the same difficulty in predicting zooplankton updates, to an even greater degree, which causes some further imbalance in the system. This becomes clear when we exclude the zooplankton types from the updating in the ML-OI (Excl. Zoo) approach (purple), since it generally equals or makes small improvements over the ML-EtE (ALL) and ML-OI (ALL) schemes.

In summary the experiments from Fig.~\ref{fig:full_variable_extension} show that the impact of DA analysis updates on the model forecast is not straightforward due to the highly non-linear and complex nature of the BGC model. Although in general it is true that increasing the number of updated variables benefits the forecasts, this is definitely not true for every variable. Furthermore, even if forecasts are improved during most of the 7-day period relative to the RUS model (as can be anticipated based on Fig.~\ref{fig:lead_time}), around the 7-day lead-time it does not really outperform the RUS approach (based on the forecast RMSEs in Fig.~\ref{fig:full_variable_extension}).

\begin{figure}[H]
    \centering
    \includegraphics[width=0.8\textwidth]{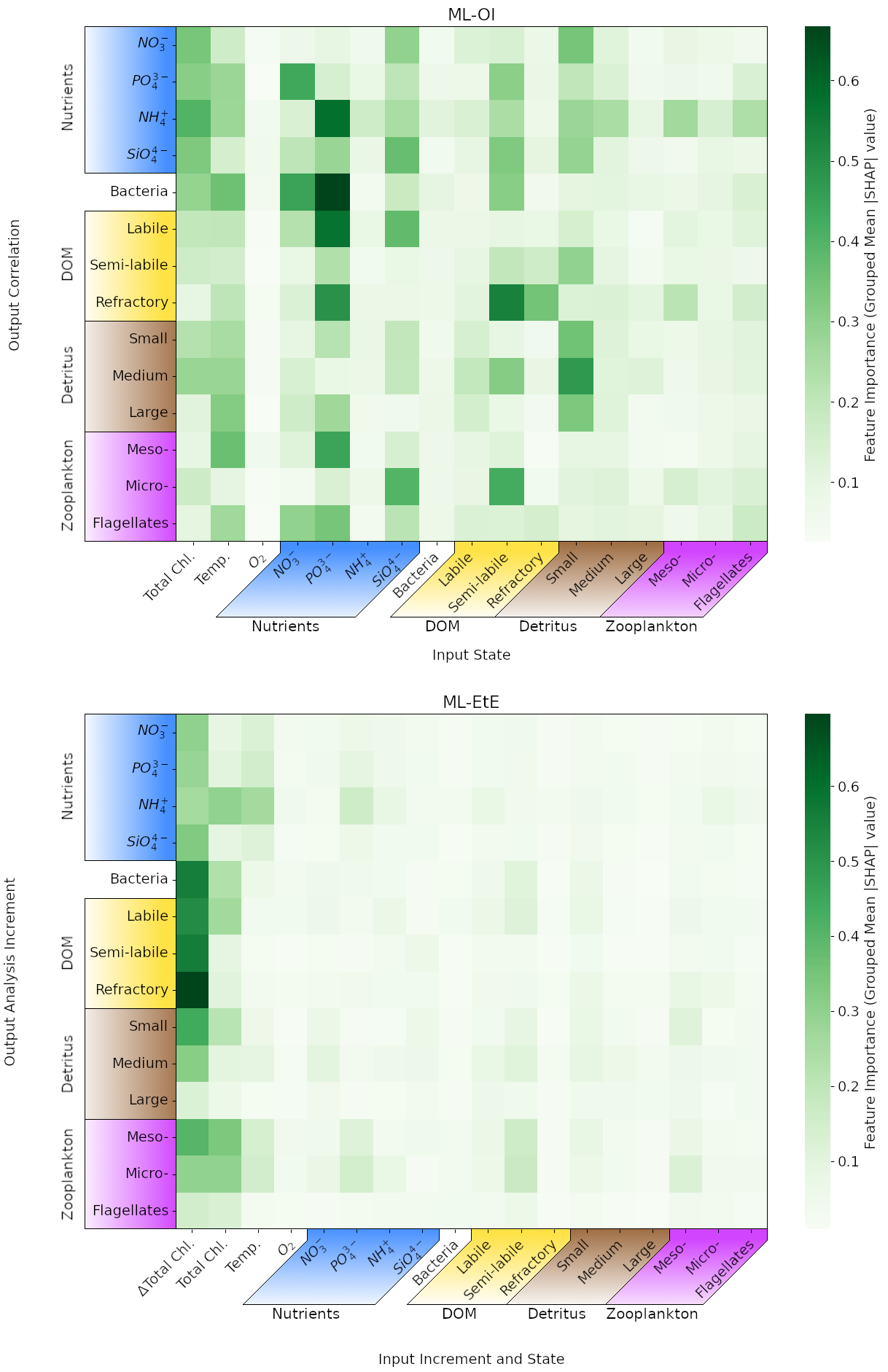}   
    \caption{Mean absolute Shapley values estimated for the ML-OI (ALL) and ML-EtE (ALL), across their respective training datasets (see Sect.~\ref{sec:new_schemes}). Chemical components that belong to the same class or type (e.g., the carbon, nitrogen and phosphorus of bacteria) have been grouped as they are highly correlated. The variable names and chemical components are detailed in Table~\ref{tab:ref_table}.
    The upper panel shows values for the extended ML-OI (ALL) model, and the lower panel for the extended ML-EtE (ALL) model. 
    The grouped input features for each ML model are given on the $x$-axis, which make up the surface state for each pelagic variable. ML-EtE (ALL) has one additional feature, $\Delta\textrm{Total Chl.}$, which represents the total chlorophyll analysis increment. 
    The grouped output targets for each ML model are given on the $y$-axis, which correspond to the correlations for ML-OI (ALL), and the analysis increments for ML-ETE (ALL).
    }
    \label{fig:shap_combined}
\end{figure}

In Fig.~\ref{fig:shap_combined}, we interrogate the ML models using Shapley values (Sect.~\ref{SHAP_metric}) to identify important ML-model features that are key to making accurate predictions, and drive the connections between observed total chlorophyll and unobserved variables. Specifically, Fig.~\ref{fig:shap_combined} shows the grouped mean absolute Shapley values for both the extended ML-OI (upper panel) and ML-EtE (lower panel) approaches. These are grouped as the separate chemical components of any class/type, and the resulting Shapley values, are very highly correlated. It is also important to note that Shapley values differ from a pure correlation between the input and output variables. This is because they capture both direct and interaction effects, account for non-linear relationships, and can explain a model’s decision-making rather than just measuring statistical association.

The ML-OI (ALL) Shapely values indicate that a broad range of input variables are important to the prediction of total chlorophyll correlations with unobserved variables, and highlight the general complexity of these interactions.
We see that the state of temperature and total chlorophyll are moderately important across a broad set of variable groups. This makes sense as this ML model is predicting the correlation of a given variable with total chlorophyll, which is generally dependent on the state of total chlorophyll. However, this also implies that the seasonal regimes play a significant role in the predictions, as temperature is a clear identifier for the current time in the seasonal cycle. 
We note that the seasonal signal of other variables could also be important for the prediction of correlations as, in some cases, we see that at least one of the state variables in a group can be important to predicting the correlation between total chlorophyll and a state variable of the same group. For example, the state of small detritus is highly important to the entire group of detritus correlations, the states of some nutrients are generally important to the prediction of nutrients, and the semi-labile DOM is somewhat important to the wider DOM correlations. 
Some input features show no strong importance to any output targets. In particular, the zooplankton types seem largely unimportant in predicting their own correlation with total chlorophyll and the variables with a stronger signal have no obvious direct relationship. This may partially explain why zooplankton performs poorly when they are updated by the ML-DA schemes, as seen previously in Fig.~\ref{fig:full_variable_extension}, and points towards the difficulty and uncertainty associated with zooplankton in marine BGC modelling. 
This is further evidenced as the zooplankton types are unimportant as input features for all other correlation predictions as well. 
We also see that oxygen is largely unimportant to the prediction of the correlations. This observation is consistent with the known weak impact of oxygen assimilation in ERSEM on other modelled variables \citep{skakala2021towards}. 
This would imply that both zooplankton and oxygen could be removed from the input feature set with little impact on the overall model performance. 

The ML-EtE (ALL) Shapley values take on a distinctly different structure to those of the ML-OI (ALL).
Recall that ML-EtE (ALL) has a different target than ML-OI (ALL), as it emulates analysis increments directly. It also has an additional input feature, the analysis increment of total chlorophyll, which is readily available in both the training dataset and at run-time. 
The most striking difference is that the total chlorophyll analysis increment dominates the prediction importances, showing the highest mean absolute value in almost all predictions. This is to be expected, as the total chlorophyll increment contains information about the observation, observational error and background model covariance, which are all necessary components of the unobserved analysis increment as described in Eq.~\eqref{eq:single_model_update}. This makes sense considering the seasonal variation of the model and that total chlorophyll represents this variation quite reliably according to the regimes discussed in Sect.~\ref{sec:sys_dynamics}.
The state variable input features show much less importance in ML-EtE (ALL) than in ML-OI (ALL), but it is sometimes still non-zero. These non-zero values seem to correlate somewhat with the most important input features seen in the ML-OI (ALL) approach, even if they are significantly reduced overall, suggesting that the state still contributes to the inherent flow dependencies of the analysis increments.

\subsection{Generalisation of machine learned-correlations to an unseen location}
\label{sec:result_unseen_location}
\begin{figure}[H]
    \centering
    \includegraphics[width=0.99\textwidth]{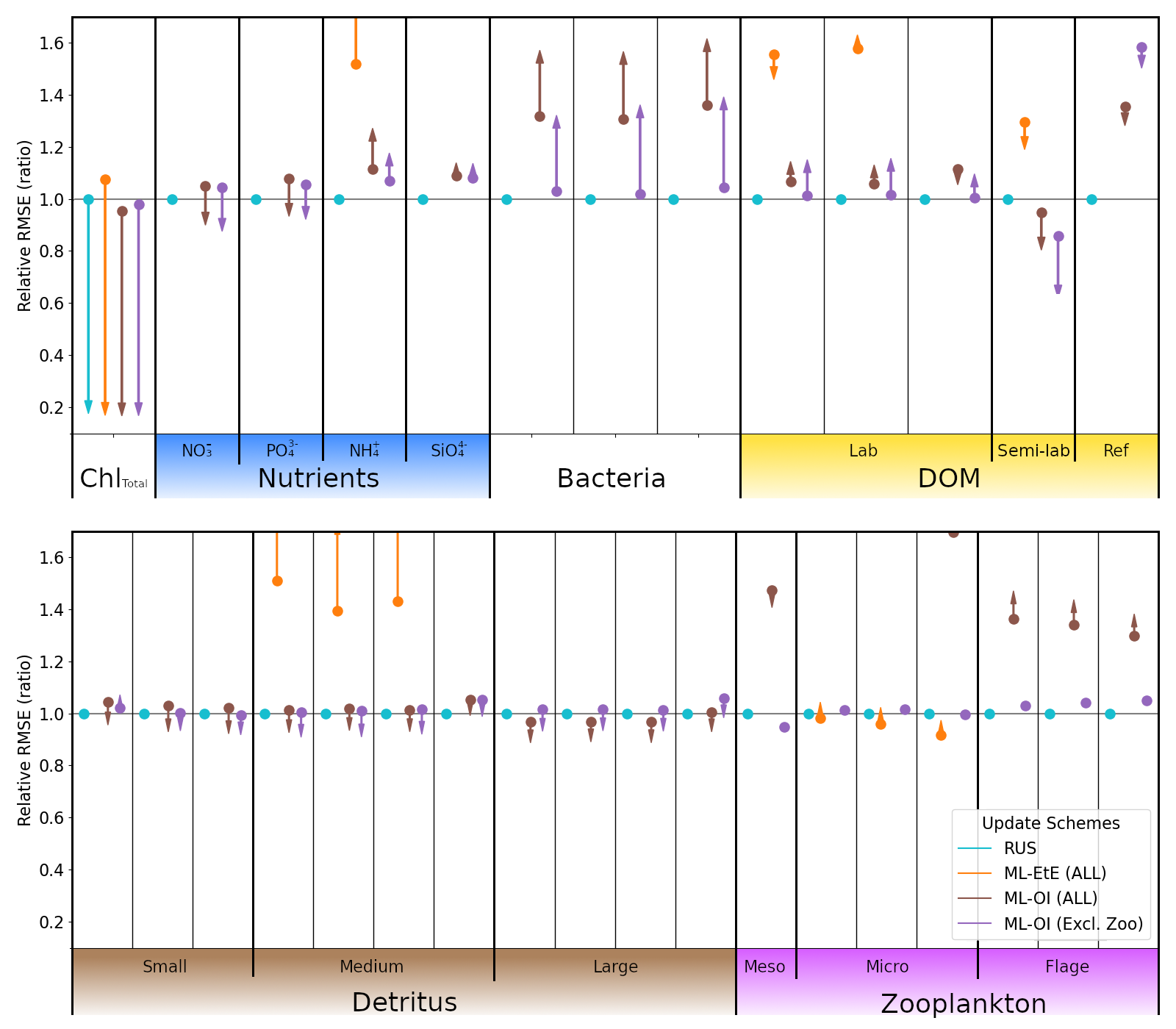}   
    \caption{As Fig.~\ref{fig:full_variable_extension}, with the ML methods trained on L4, but applied to the new location CWEC. Dots and arrows that do not appear are off the scale.
    }
    \label{fig:EC_conditions}
\end{figure}

\begin{figure}[H]
    \centering
    \includegraphics[width=0.99\textwidth]{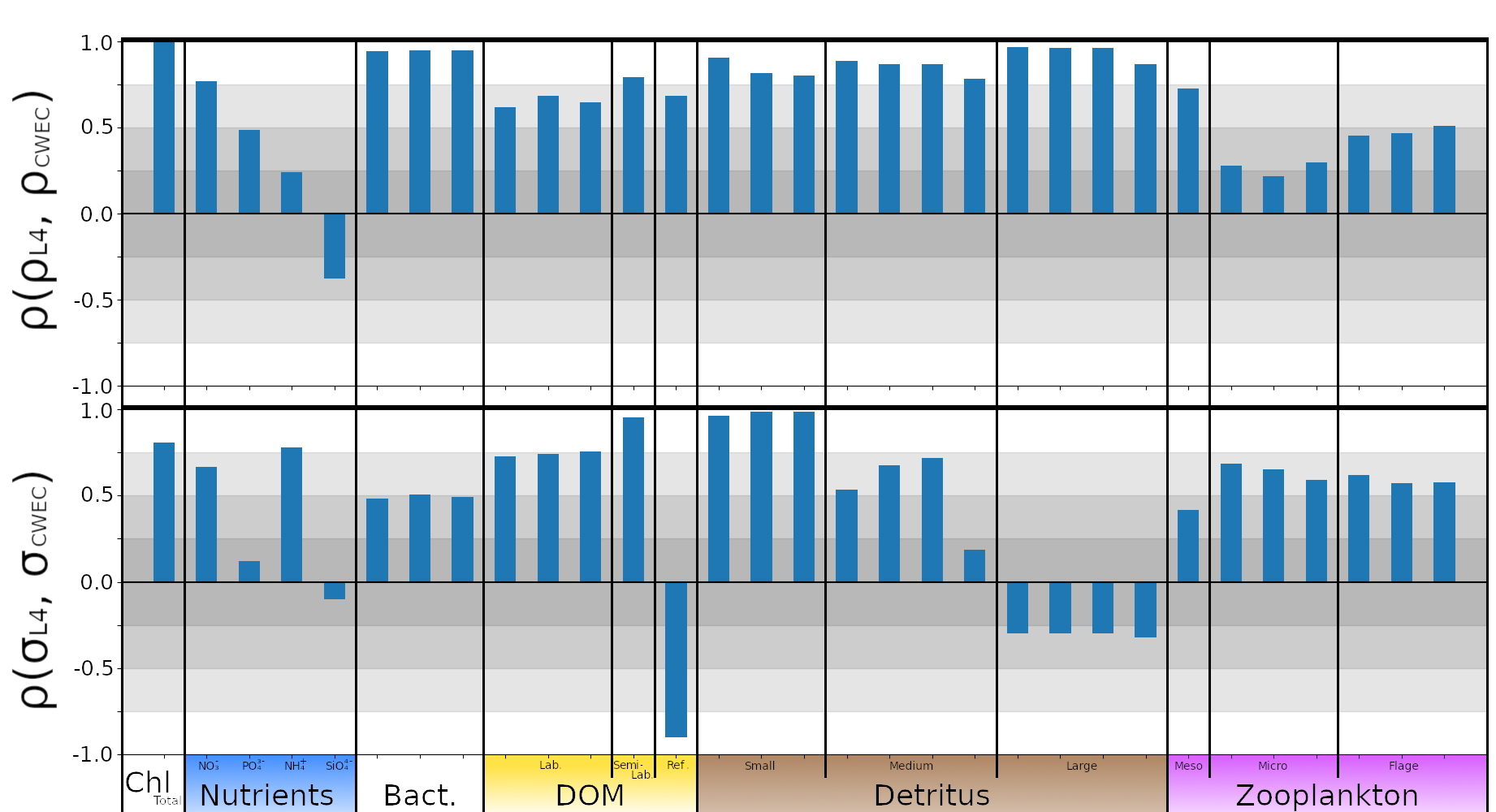}   
    \caption{The top panel shows the correlation between a given variable's climatological correlation signal at L4 and the CWEC, $\rho(\rho_{L4}, \rho_{CWEC})$. 
    The bottom panel shows the correlation between a given variable's climatological standard deviation signal at L4 and the CWEC, $\rho(\sigma_{L4}, \sigma_{CWEC})$.
    High correlations indicate that the model is behaving similarly in each location.}
    \label{fig:location_similarity}
\end{figure}

In this section, we test the performance of the extended ML approaches from Sect.~\ref{sec:result_extended_variable_set} in the CWEC (Fig.~\ref{fig:location_map}), which exhibits different marine BGC behaviour than the L4 training location. In Fig.~\ref{fig:EC_conditions}, we assess the performance of these ML models according to their 7-day forecast and analysis RMSEs. We then compare some general differences between the climatology of the two locations in Fig.~\ref{fig:location_similarity}, and then, with reference to the Shapley values shown previously in Fig.~\ref{fig:shap_combined}, we shall discern why the ML model might struggle extrapolating to the new location.

Figure~\ref{fig:EC_conditions} again uses the RUS scheme as a comparison point for the ML model approaches, so all RMSEs are given as a ratio of the RUS's RMSE value at the new location.
The ML-EtE (ALL) approach (orange) performs extremely poorly in this new location, with a large portion of the RMSEs exceeding $1.5\times$ the RUS background error (off the scale of Fig.~\ref{fig:EC_conditions}). This is because the emulated analysis increments of the EnKF at the L4 location fit the variability and scale of that (trained) location and so, do not translate well to the new location. This means that, while the ML-EtE (ALL) approach works well at the trained location (and fits the expected distribution of input data), in practice its extendability to a new location is limited by both availability of training data and to the new location's similarity to the original training location.
The ML-OI (ALL) (brown) makes a marked improvement over the ML-EtE (ALL) scheme, which is the reverse of the previous scenario at the L4 location. This is likely because the correlations predicted by the ML-OI scheme represent a more location-agnostic relationship in the marine BGC variables, which can be used in combination with the climatological variances of CWEC to produce more location-appropriate increments. However, this scheme still struggles to predict zooplankton correlations, and so not updating the zooplankton as in the ML-OI (Excl. Zoo) scheme (purple) produces a broadly improved result. 
In both ML-OI (ALL) and ML-OI (Excl. Zoo), we see that detritus is generally improved relative to the RUS scheme. Figure~\ref{fig:location_similarity} shows that the climatological correlations for these variables are generally similar in both locations (compare Fig.~\ref{fig:app:climatology_corrs_L4} and Fig.~\ref{fig:app:EC_climatology_corrs} to see how these vary with time), with small detritus (originating largely from species with size $<20\mu m$) showing similarity in both climatological correlation and standard deviation. Since small detritus is the most important input feature, in Fig.~\ref{fig:shap_combined}, for the prediction of detritus correlations in the ML-OI models, it is reasonable to see why the improvements persist between the two locations. 
We also see in Fig.~\ref{fig:EC_conditions} that both ML-OI models (brown and purple) make improvements to the analysis RMSEs of nitrate, phosphorus and semi-labile DOM, which show relatively similar climatological behaviour to L4 in Fig.~\ref{fig:location_similarity} and Fig.~\ref{fig:app:behaviour_dist}.

As all training is performed at one location, it is easy to hypothesise that the ML models have overfitted to L4, specifically with regards to their use at other locations. Here, L4 is coastal and the CWEC a more open area of ocean. This does not rule out the possibility that ML models trained on a limited number of locations could extend their predictions to spatial locations beyond their set of training locations. However, it indicates that sparse training locations would need to be chosen carefully, to appropriately cover the spread of behaviour in the system.

\subsection{Viewpoint on scaling multivariate data assimilation to 3D models}
\label{sec:3d}
We have successfully shown that ML methods can make improvements to the DA schemes of marine BGC models when coupled to a 1D physical model. The natural next question is how these results would scale when the marine BGC model is coupled to a 3D physical model, such as NEMO (Nucleus for European Modelling of the Ocean).
A seemingly simple solution would be to run (once) a well-tuned, large EnKF, which can then be used to train an ML model to be used operationally in an analogous 3D DA system that presently updates only total chlorophyll. 
A reanalysis product with comprehensive statistics, or all ensemble members available so statistics can be generated, would be ideal \citep{bonavita2020machine, brajard2021combining, gregory2024machine}. This circumvents the need to run an expensive DA scheme operationally as the ML model could be trained offline, and then run significantly faster while retaining the benefit of the statistics learned from a large ensemble. This would also allow the analysis increments to be predicted directly. However, state-of-the-art ensemble marine BGC systems are still limited in scale and may not (yet) accurately represent the statistics needed for multivariate DA \citep{skakala2024uncertain}. Also, this approach would need to be repeated if/when the observation network changes, which is likely given new observation missions and strategies \citep{telszewski2018biogeochemical}.
A cheaper alternative would be to calculate the correlations in a free-run ensemble dataset, as per the methods described in Sect.~\ref{sec:method_mlfc}. This would be cheaper to create as there would be no need to store and calculate both background states and analysis states. However, this approach cannot calculate the analysis increments directly and instead must rely on the hybridisation of background covariances/correlations into existing DA frameworks. Nevertheless our results on the 1D scenario suggests that this is feasible and a good alternative to estimating the increments directly.

It is also worth considering how the data for these ML models should be sampled spatially in the 3D case. Our results show there is some transferability between locations, as long as the dynamics are similar enough. In this, we suggest that a sparse forest of 1D models could be generated across the 3D domain, which aims to cover each region of sufficiently different biogeochemical behaviour. Previous work by \cite{higgs2024investigating} has split the North-West European Shelf into dynamically connected ecoregions, and this, or similar analysis, could be used as a guideline for generating these 1D models.
A limitation of our two test locations is that they are not directly coupled, and could only be considered weakly coupled in the sense that their forcing data is extracted from the same 3D weather model. This could mean that 3D models have an advantage in locations having similar behaviour, as model grid points are much more likely to strongly correlate due to advection and ocean currents. However, the inverse could also be true, as the 1D models do not consider riverine input which can have substantial effects at the coast. Either way, the results suggest that some sparsity could be applied in extracting training data for these models, as long as each regime of BGC behaviour is represented in the selection. Introducing spatial variables like longitude and latitude could also improve the models ability to predict increments or correlations across the different horizontal locations.

\section{Conclusions}
Marine biogeochemistry (BGC) models aim to represent the complex BGC processes necessary to understand and forecast ecosystem behaviour. Data assimilation (DA) plays a crucial role in ensuring model trajectories remain closely aligned with real-world observations, along with the need for continuous improvement of numerical predictions. Both numerical modelling and DA are computationally expensive for marine BGC (dealing with great complexity and many variables), requiring well-tuned and accurately sampled statistics to be effective. These statistics are often poorly estimated in the undersized ensemble-based methods that are affordable operationally. In turn, this leads to the use of climatological forecast error covariance matrices in deterministic models, or simply not updating unobserved variables. This section concludes our work in relation to the research questions set out towards the end of Sect.~\ref{sec:background} (reproduced below in italics).

\textit{(a) Can we make improvements to the existing univariate scheme by updating a limited set of additional variables with an ML model to predict correlations or analysis increments?} In this study, we have demonstrated that neural networks can effectively learn statistical relationships between total chlorophyll (the only observed variable) and various pelagic BGC model variables. With machine learning (ML), we achieve significant improvements over conventional approaches that rely on climatological statistics or omit updates altogether. Our analysis of ML-predicted nitrate updates illustrates that the ML methods behave in a largely coherent and meaningful manner, reinforcing their potential as an effective tool for improving BGC DA.

\textit{(b) Can these ML models be extended to effectively update all unobserved pelagic variables?} ML models can update almost all unobserved pelagic variables, supporting the broader applicability of ML in DA. Some variables (notably zooplankton) do not update well using either of the ML methods that are extended to update all state variables (namely ML-OI (ALL) and ML-EtE (ALL)). 
Zooplankton variables are better treated in these hybrid DA schemes without being updated directly (as in ML-OI (Excl. Zoo)).

\textit{(c) Is the ML model transferable to a new location after being trained on some other location?} While a neural network trained in one water column exhibits partial transferability to other locations, challenges remain in fully generalising the model across spatial domains. This partial transferability is valuable, given the difficulty and cost of acquiring high-quality training data across large oceanic regions, and should be explored further in the context of 3D models. We discuss the feasibility of this, and propose a methodology for doing so. Future work should focus on refining transferability strategies, effective sampling strategy to allow for ergodic coverage and further evaluating the scalability of ML-driven DA in complex marine environments.

\newpage
\appendix

\setcounter{figure}{0} 
\counterwithin{figure}{section}
\section{Characterisation of location biogeochemistry}
\label{app:a}

\begin{figure}[H]
    \centering
    \includegraphics[width=0.99\textwidth]{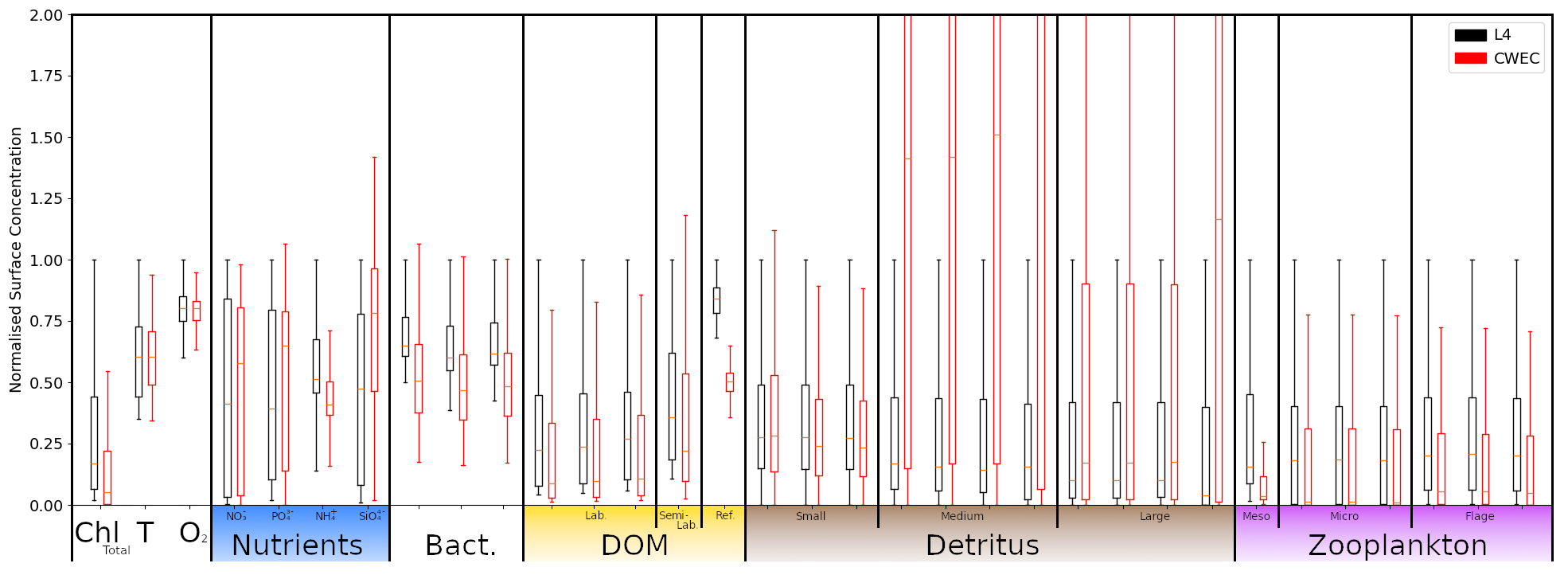}   
    \caption{Box and whisker plot showing the $25^{th}$, $50^{th}$ and $75^{th}$ percentile and upper and lower bound (excluding outliers larger than $1.5\times\textrm{Inter quartile range}$) of each pelagic marine BGC variable for the RUS scheme in the online testing period. Values for L4 are given in black, and values for the CWEC are given in red. All values are normalised against the upper bound of L4. Chemical components are ordered according to Table~\ref{tab:ref_table}. The label ``T'' corresponds to temperature.
}
    \label{fig:app:behaviour_dist}
\end{figure}

Figure~\ref{fig:app:behaviour_dist} shows that the CEWC is clearly less biologically productive than L4  with surface concentrations of total chlorophyll having a significantly lower median value, and a maximum that is approximately 50\% of L4's maximum.
Each exhibit similar temperature values, as they are both located within the English Channel. 
In the nutrients group, nitrate and phosphate values cover a similar range in each location, but ammonium and silicate have little overlap.
Bacteria and DOM concentrations also show little similarity between locations.
The small detritus concentrations are very similar between both locations, but the medium and large detritus differ significantly, with CWEC covering a much wider range of values than L4.
Zooplankton concentrations also differ between the locations, with CWEC producing much lower concentrations of zooplankton than the more biologically active L4 location.

\begin{figure}[H]
    \centering
    \includegraphics[width=0.99\textwidth]{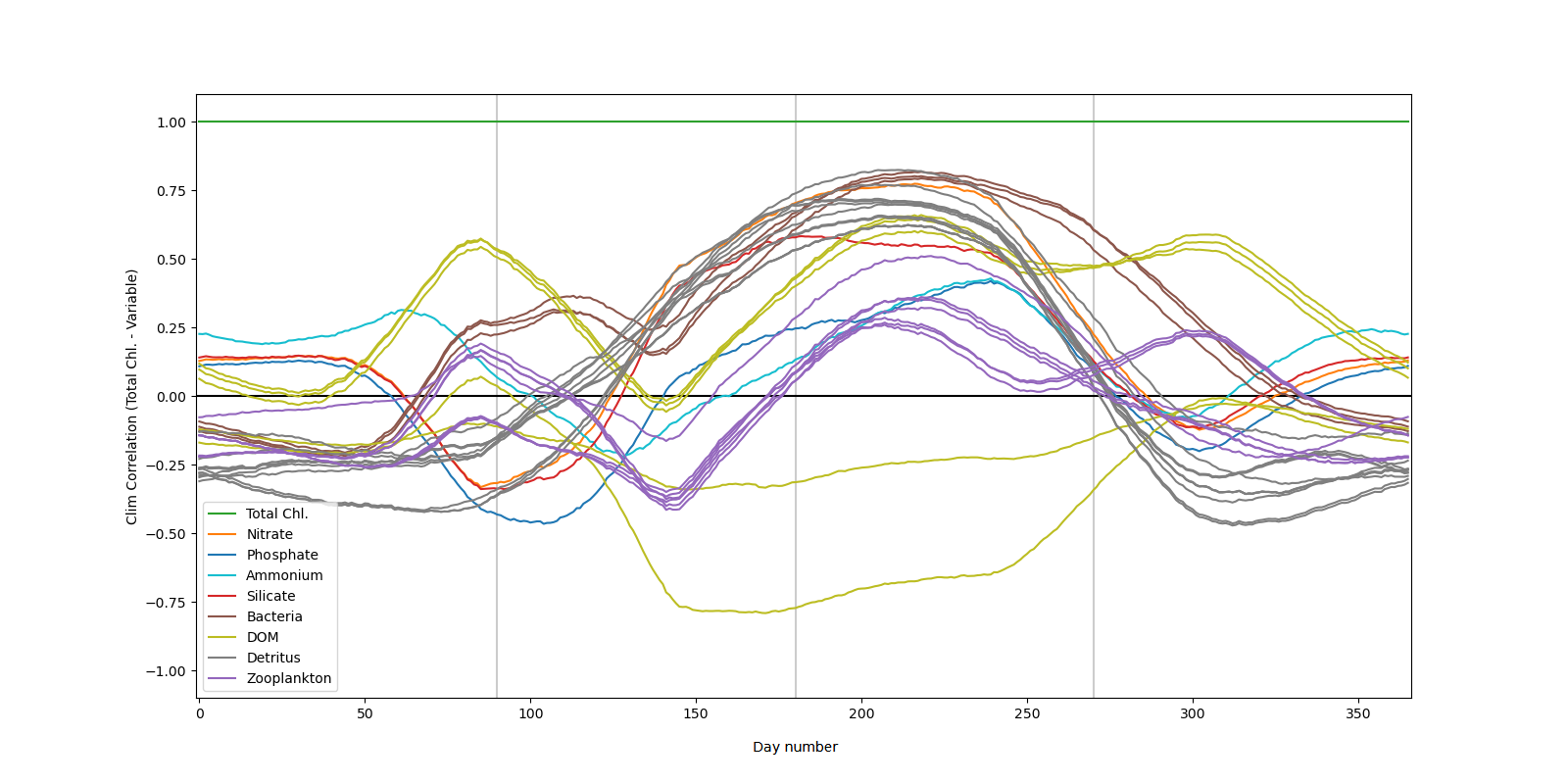}   
    \caption{Daily climatological correlations for each pelagic variable at the L4 location, calculated from the training free-run period of 2000-2014. Pelagic variables of the same class (according to Table~\ref{tab:run_types}) are shown with the same colour, except nutrients (nitrate, phosphate, ammonium and silicate) which are shown with separate colours.
}
    \label{fig:app:climatology_corrs_L4}
\end{figure}

The climatological correlations between total chlorophyll and other pelagic variables at the L4 location, shown in Fig.~\ref{fig:app:climatology_corrs_L4}, vary significantly according to the season. 
Variables of the same class (see Table~\ref{tab:ref_table}) generally exhibit very similar correlations.
Correlations are much stronger during the spring and summer months, as this period is more biologically active, and so the different model components are going to be more closely coupled. Some variables, such as zooplankton, show a much weaker correlative relationship with total chlorophyll.

\begin{figure}[H]
    \centering
    \includegraphics[width=0.99\textwidth]{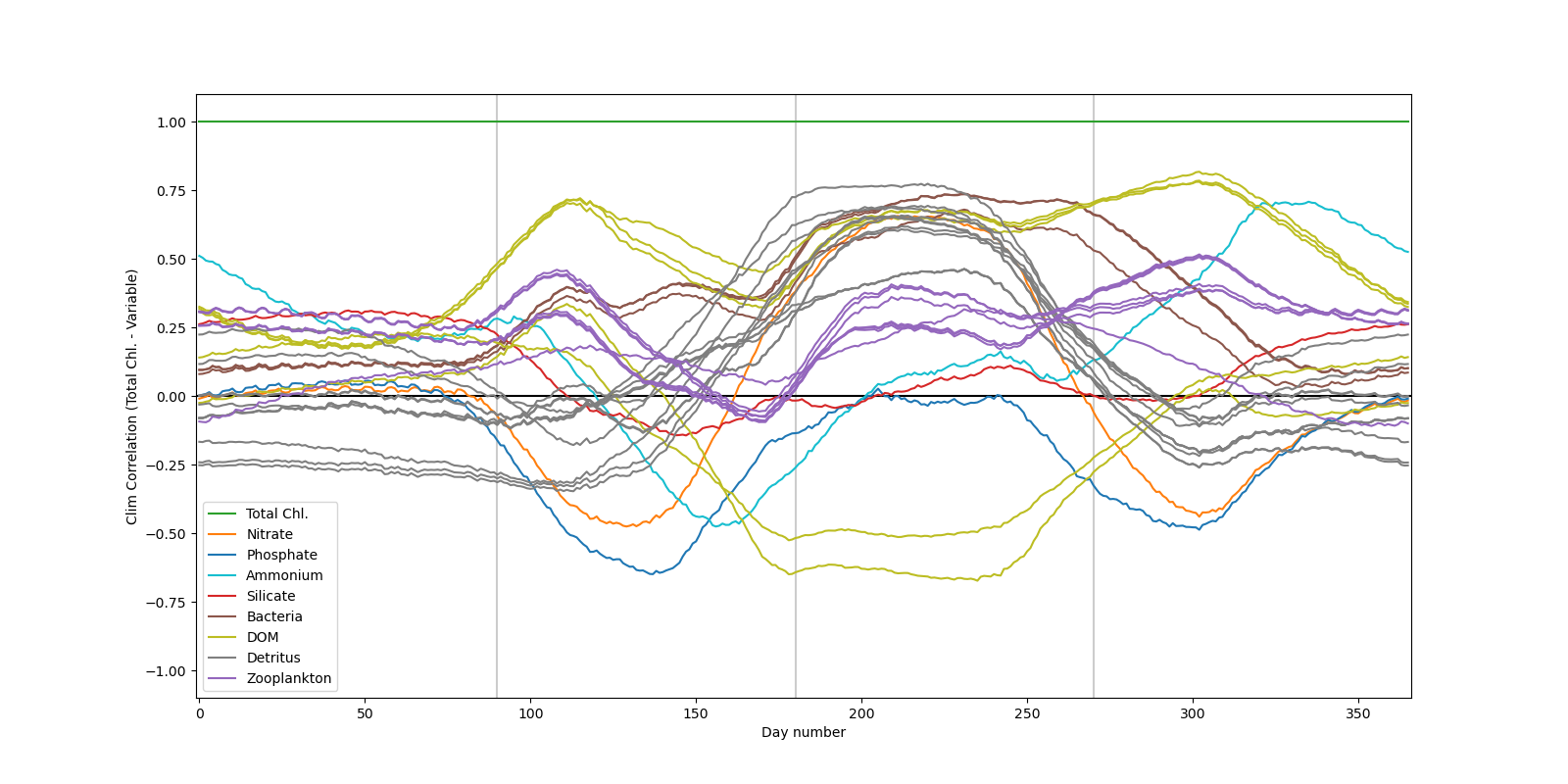}   
    \caption{As with Fig.\ref{fig:app:climatology_corrs_L4}, except for the CWEC location from a free-run period of 2000-2010.
}
    \label{fig:app:EC_climatology_corrs}
\end{figure}

Figure~\ref{fig:app:EC_climatology_corrs} shows the climatological correlations between total chlorophyll and other pelagic variables at the CWEC location. 
As with L4, the correlations of most variables show a much stronger correlation with total chlorophyll during the spring and summer, when the system is much more active.
The correlations of nitrate are similar to those seen at the L4 location in Fig.~\ref{fig:app:climatology_corrs_L4}, following the pattern described Sect.~\ref{sec:sys_dynamics}. Zooplankton shows a weak correlation with total chlorophyll.

\newpage
\bibliography{bibliography.bib}

\end{document}